\documentclass[aps,prl,twocolumn,preprintnumbers,amsmath,amssymb,superscriptaddress,10pt]{revtex4-2}
\usepackage{amsmath,graphicx}
\usepackage[utf8]{inputenc}
\usepackage[T1]{fontenc}
\usepackage{xcolor}
\usepackage{textcomp}
\usepackage{bm}
\usepackage{slashed}
\usepackage{physics}
\usepackage{amsmath} 
\usepackage{tikz}
\usepackage{mathdots}
\usepackage{yhmath}
\usepackage{cancel}
\usepackage{color}
\usepackage{array}
\usepackage{multirow}
\usepackage{amssymb}
\usepackage{gensymb}
\usepackage{tabularx}
\usepackage{extarrows} 
\usepackage{booktabs}
\usetikzlibrary{fadings}
\usetikzlibrary{patterns}
\usetikzlibrary{shadows.blur}
\usetikzlibrary{shapes}

\usepackage{color}
\definecolor{LinkColor}{rgb}{0.75,0.0,0.2}

\usepackage{hyperref}
\hypersetup{
	pdfauthor={good guys},
	pdftitle={good title},
	colorlinks=true,
	citecolor=LinkColor,
	linkcolor=LinkColor,
	urlcolor=LinkColor,
}

\usepackage{listings}
\definecolor{lightgray}{gray}{1}

\usepackage{theorem}

\usepackage{tikz}

\newcommand{\nc}{\newcommand}
\newcommand{\Rmnum}[1]{\uppercase\expandafter{\romannumeral#1}}
\nc{\braoprket}[3]{\langle#1|#2|#3\rangle}
\nc{\opn}[1]{\operatorname{#1}}
\nc{\avg}[1]{\langle#1\rangle}
\nc{\ketbrasame}[1]{|#1\rangle\!\langle#1|}
\nc{\swap}{\opn{SWAP}}
\nc{\E}{\mathbb{E}}
\nc{\Var}{\opn{Var}}
\nc{\dg}{\dagger}

\usepackage[normalem]{ulem}

\begin{document}

\title{Anomalous Dynamical Scaling at Topological Quantum Criticality}
\author{Menghua Deng}
\affiliation{School of Physics and Electronics, Hunan University, Changsha 410082, China}
\author{Sheng Yang}
\affiliation{Institute for Advanced Study in Physics and School of Physics, Zhejiang University, Hangzhou 310058, China}
\author{Chen Sun}
\affiliation{School of Physics and Electronics, Hunan University, Changsha 410082, China}
\author{Fuxiang Li}
\email{fuxiangli@hnu.edu.cn}
	\affiliation{School of Physics and Electronics, Hunan University, Changsha 410082, China}
    
\author{Xue-Jia Yu}
\email{xuejiayu@eitech.edu.cn}
\affiliation{Eastern Institute of Technology, Ningbo 315200, China}

\begin{abstract}
We study the nonequilibrium driven dynamics at topologically nontrivial quantum critical points (QCPs), and find that topological edge modes at criticality give rise to anomalous dynamical scaling behavior.  By analyzing the driven dynamics of bulk and boundary order parameters at topologically distinct QCPs in quantum spin chains, we demonstrate that, while the bulk dynamics remain indistinguishable and follow standard  Kibble–Zurek (KZ) scaling,  the anomalous boundary dynamics are unique to topological criticality, obeying 
modified scaling relation beyond the traditional KZ framework. To elucidate the unified origin of this anomaly, we further study the dynamics of defect production at topologically distinct QCPs in free-fermion models and demonstrate similar anomalous scaling exclusive to topological criticality. These findings establish the existence of anomalous dynamical scaling arising from the interplay between topology and driven dynamics, challenging standard paradigms of quantum critical dynamics.


\end{abstract}

\maketitle

\emph{Introduction.}---Non-equilibrium dynamics constitutes a central facet of quantum phase transitions and plays a crucial role in understanding quantum critical phenomena. A celebrated example is the Kibble–Zurek (KZ) mechanism, originally proposed in cosmology~\cite{kibble1976IOP,KIBBLE1980183,kibble2007AIP} and later applied broadly in condensed matter physics~\cite{Zurek1985Nature,ZUREK1996177,Damski2005PRL,Zurek2005PRL,Polkovnikov2005PRB,Dziarmaga2005PRL,Sen2008PRL,Barankov2008PRL,DeGrandi2010PRB,del2014IJMPA} and adiabatic quantum computing~\cite{Edward2001Science,Schutzhold2006PRA}. According to KZ mechanism, when a system is slowly quenched across a critical point at a finite rate, a nonzero density of defects is generated, exhibiting universal scaling behavior uniquely determined by the critical exponents of the phase transition~\cite{Dziarmaga2010AIP,Polkovnikov2011RMP}. Over the past decades, the KZ mechanism and its generalizations have been extensively studied from both theoretical~\cite{Bermudez2009PRL,Lee2015PRB,Dutta2016PRL,lkacki2017JOSM,Yin2017PRL,Bialonczyk2018JOSM,Liou2018PRB,del2018PRL,Dora2019NC,Ulifmmode2019PRB,Rams2019PRL,Revathy2020PRR,Bialonczyk2020IOP,Ulifmmode2020PRL,Kristor2020SciPost,Nowak2021PRB,Mayo2021PRR,Rysti2021PRL,del2022PRB,Gomez2022PRB,Sun2022PRB,Kou2023PRB,Balducci2023PRL,Zeng2023PRL,Balducci2023PRL,Xia2023PRD,Kou2022PRB,Liang2024PRB,Grabarits2025PRA,Deng2025PRL,Deng2025PRB,Kou2025PRB,wang2025NC,yang2025topologicaldefectformationkibblezurek,Zhang2025PRB_a} and experimental~\cite{Weiler2008Nature,Lamporesi2013NP,Deutschl2015PNAS,Nir2015Science,Anquez2016PRL,Ko2019NP,Keesling2019Nature,Yi2020PRL,Ebadi2021Nature,Huang2021PRL,Maegochi2022PRL,Du2023NP,Yuan2025PRL,Zhang2025PRL} perspectives and are thus widely regarded as the standard paradigm for analyzing the dynamics of systems driven across phase transitions.

On a different front, the discovery of topological physics at quantum critical points (QCPs)~\cite{YU20261} has challenged the long-standing belief that the universality class of a quantum phase transition is uniquely determined by a set of critical exponents. Specifically, topological edge modes—previously thought to exist only in gapped systems—have been shown to persist in gapless quantum critical systems, giving rise to the notion of topologically nontrivial QCPs~\cite{Verresen2018PRL,verresen2020topologyedgestatessurvive,Verresen2021PRX,Duque2021PRB,Yu2022PRL,Ye22022SciPost,Wang2022SciPost,Jones2023PRL,Mondal2023PRB,Yu2024PRB,Wei2024JHEP,Choi2024PRB,Prembabu2024PRB,Zhong2024PRA,Flores2025PRL,Zhong2025PRB,Huang2025PRB,Huang2025SciPost,Li2025PRB,Rey2025PRB,Zhou2025CP,Cardoso2025PRB,chou2025ptsymmetryenrichednonunitarycriticality,tan2025exploringnontrivialtopologyquantum}, or more generally, gapless symmetry-protected topological (gSPT) states~\cite{Scaffidi2017PRX,Parker2018PRB,JIANG2018753,Parker2019PRL,Thorngren2021PRB,Hidaka2022PRB,Wen2023PRB,Yu2024PRL,Zhang2024PRA,Su2024PRB,Li2024SciPost,ando2024gauginglatticegappedgaplesstopological,wen2024topologicalholographyfermions,Jia2025PRL,Li2025SciPost,Yang2025CP,Wen2025PRB,yu2025gaplesssymmetryprotectedtopologicalstates,yang2025deconfinedcriticalityintrinsicallygapless,bhardwaj2025gaplessphases21dnoninvertible,wen2025stringcondensationtopologicalholography,wen2025topologicalholography21dgapped,guo2025generalizedlihaldanecorrespondencecritical,prembabu2025multicriticalitypurelygaplessspt}. This conceptual extension has reshaped our understanding of quantum phase transitions, which can also exhibit nontrivial topological properties beyond gapped counterparts, including nontrivial conformal boundary conditions~\cite{Yu2022PRL,Parker2018PRB}, algebraically localized edge modes~\cite{Verresen2021PRX,Yang2025CP}, universal bulk–boundary correspondence~\cite{Yu2024PRL,Zhong2025PRB,guo2025generalizedlihaldanecorrespondencecritical}, and intrinsically gapless topological states~\cite{Thorngren2021PRB,Zhang2024PRA,Wen2023PRB}.

However, despite the extensive attention that gSPT physics has attracted in recent years, most studies have focused on equilibrium properties, and its nonequilibrium dynamics remains largely unexplored. From a fundamental perspective, can the nontrivial topology associated with QCPs induce anomalous dynamical scaling behaviors? From a practical perspective, exploring the nonequilibrium dynamics offers a realistic protocol for detecting gSPT physics in modern quantum platforms. 
These considerations motivate us to investigate the interplay between topology and nonequilibrium driven critical dynamics.

In this Letter, we uncover that nontrivial topology at criticality will lead to anomalous scaling behaviors beyond the conventional KZ framework. Specifically, we first examine the driven dynamics at prototypical topologically distinct Ising QCPs—the transverse-field Ising and cluster-Ising transitions. While the bulk dynamics at both critical points follow standard KZ scaling, the boundary dynamics differ strikingly: the transverse field Ising QCP exhibits conventional KZ behavior, whereas the cluster-Ising QCP obeys a modified power-law scaling  beyond the traditional KZ prediction. This scaling is further numerically confirmed in the non-exactly solvable quantum Potts chains. To elucidate the unified origin underlying this anomalous behavior, we further investigate driven dynamics at topologically distinct QCPs in free-fermion models and observe similar anomalous scaling in edge excitations at topologically nontrivial QCPs. These results unambiguously demonstrate that robust topological edge modes at criticality are the essential ingredient driving anomalous universal dynamical behaviors, thus establishing a new mechanism for observing dynamical phenomena beyond the KZ framework.

\begin{figure}[t]
	\centering
\includegraphics[width=3.5in]{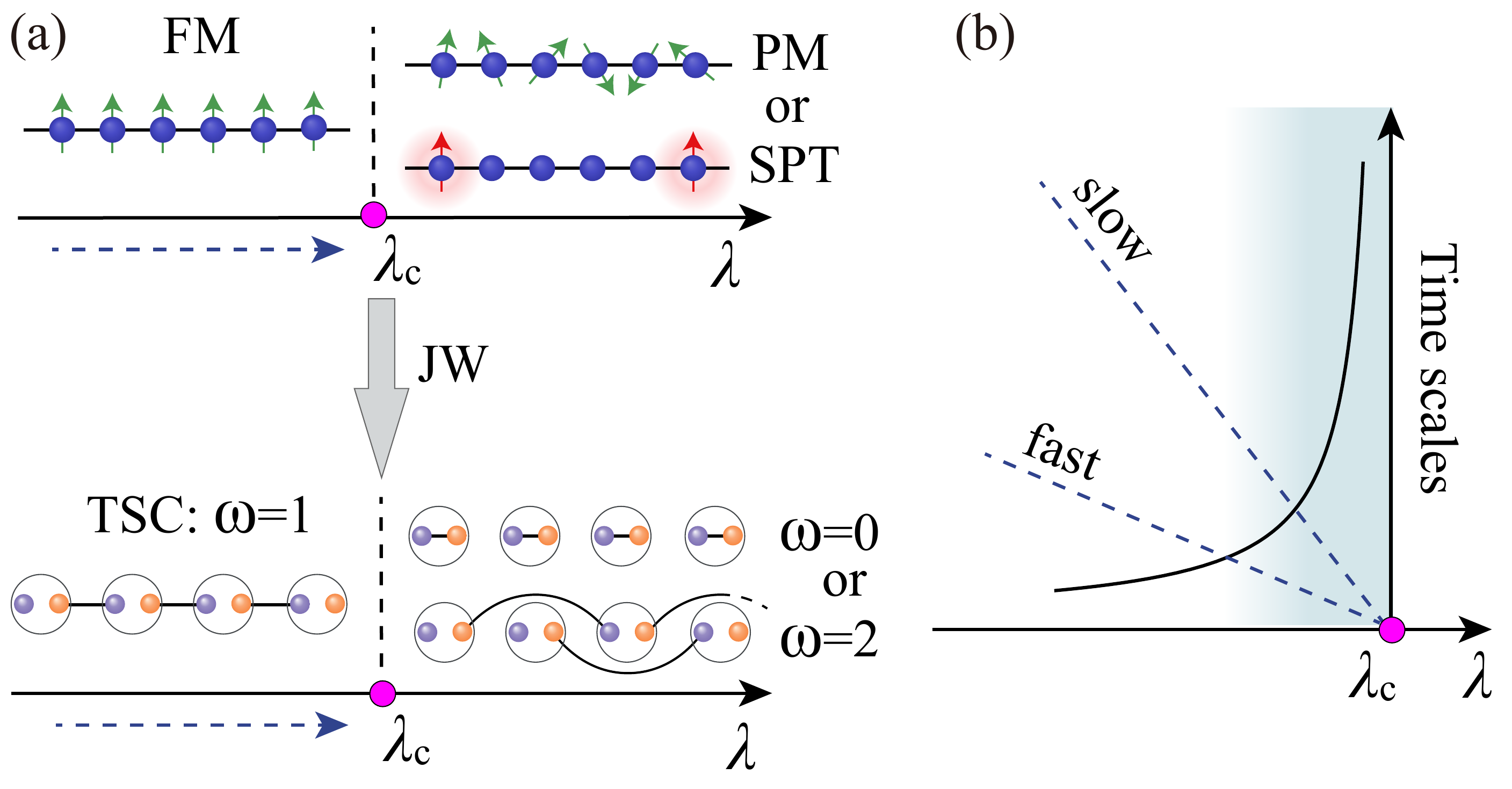}\\
	\caption{(a) Schematic phase diagram of the spin model in Eq.~(\ref{E1}) (upper panel) and its fermionic dual (lower panel). In the spin models, the control parameter $\lambda$  tunes the system across the topologically nontrivial or trivial QCP $\lambda_c$, separating the FM phase  from the cluster-SPT phase  or the PM phase. In the fermion model, the control parameter $\lambda$  tunes the system across the topologically nontrivial or trivial QCP $\lambda_c$, separating the topological superconductor (TSC) phase  with winding number $\omega = 1$   from the TSC with $\omega = 2$ or topologically trivial with $\omega = 0$. The control parameter $\lambda$ is linearly ramped from the FM phase (TSC with $\omega = 1$) toward the QCP at $\lambda_c$ (blue dashed arrow). 
    (b) Relevant time scales for driven dynamics near criticality. The correlation time scales diverges as the system approaches the QCP (black solid curve), while the blue dashed line denotes the time distance $|\lambda - \lambda_c|/R$ to the QCP for different quench rates.} \label{fig:1}
\end{figure}

\emph{Warm-up: Quantum Ising chains.}---To illustrate the gSPT physics more concretely, we consider a transverse‐field spin-1/2 chain with cluster interactions~\cite{Verresen2021PRX,Yu2022PRL,Yu2024PRB}:
\begin{equation}
\label{E1}
H_{\text{spin}} =-\sum_{i}J \sigma^{z}_{i}\sigma^{z}_{i+1}-\sum_{i} h \sigma^{x}_{i} -\sum_{i} J_3 \sigma^{z}_{i-1}\sigma^{x}_{i}\sigma^{z}_{i+1},
\end{equation}
The model possesses a $\mathbb{Z}_{2}$ spin-flip symmetry generated by
$P=\prod_{i}\sigma^{x}_{i}$ and a time-reversal symmetry $\mathbb{Z}_{2}^{T}$ generated by complex conjugation $T=K$, where $T$ denotes the antiunitary operator that generate time-reversal symmetry and $K$ represents complex conjugation. Here, $J$ denotes the Ising coupling (set as the energy unit), while $h$ and $J_3$ control the transverse field and the cluster interaction, respectively. For $J_3=0$, the Hamiltonian reduces to the transverse-field Ising (TFI) model, which exhibits a continuous transition between the ferromagnetic (FM) and paramagnetic (PM) phases, while for $h=0$, the Hamiltonian reduces to the cluster–Ising (CI) model, which exhibits a continuous transition between the FM and cluster SPT phases, as shown in Fig.~\ref{fig:1} (a). More importantly, although the FM–PM and FM–SPT transitions share the same critical exponents and are expected to belong to the (1+1)-dimensional Ising universality class~\cite{Yu2024PRB,Duque2021PRB}, they are nevertheless distinguished by the topological properties: the latter hosts robust topological edge modes even at criticality. This distinction leads to the notion of topologically nontrivial QCPs (or gSPT states), referred to as the Ising and $\text{Ising}^{*}$ QCPs for the TFI and CI chains, respectively. The central theme of this work is to uncover the novel dynamical scaling behavior that arises at such topological quantum criticality.

\emph{Anomalous  dynamical  scaling of boundary magnetization}---Turning to the dynamics at the Ising and $\text{Ising}^{*}$ QCPs in the interacting spin chain of Eq.~(\ref{E1}), we examine the dynamical scaling of the local magnetization at site $l$, defined as~\cite{Yang1952PR,Igloi2011PRL} $m_l(t)=\langle \Psi(t) | \sigma_l^{z} | \Psi(t) \rangle$, where $|\Psi(t)\rangle$ is the time-evolved state vector. We study the  driven critical dynamics of the TFI and CI models by linearly ramping the control parameters $\lambda = h, J_3$ according to \begin{equation}
\lambda(t) = \lambda_c +( \lambda_c-\lambda_i) R t, ~~t \in [-1/R, 0],
\end{equation}
which drives the system from the FM phase to the critical point $\lambda_c$ (see Fig.~\ref{fig:1} (b)). We analyze the resulting dynamical scaling of both the bulk $m_{L/2} \equiv m_b$ and the boundary magnetization $m_s$~\footnote{For the topologically trivial QCPs, the boundary spin is chosen as the first site, while for the topologically nontrivial QCPs, the boundary spin is chosen as the second site. More details on this subtlety can be found in Sec.~II of the Supplemental Material.} in systems of size $L$ with open boundary conditions. Technical details of the Gaussian-state method used in these calculations are provided in Sec.~\Rmnum{1} of the Supplemental Material(SM)~\cite{sm}.

\begin{figure}[t]
	\centering
\includegraphics[width=3.5in]{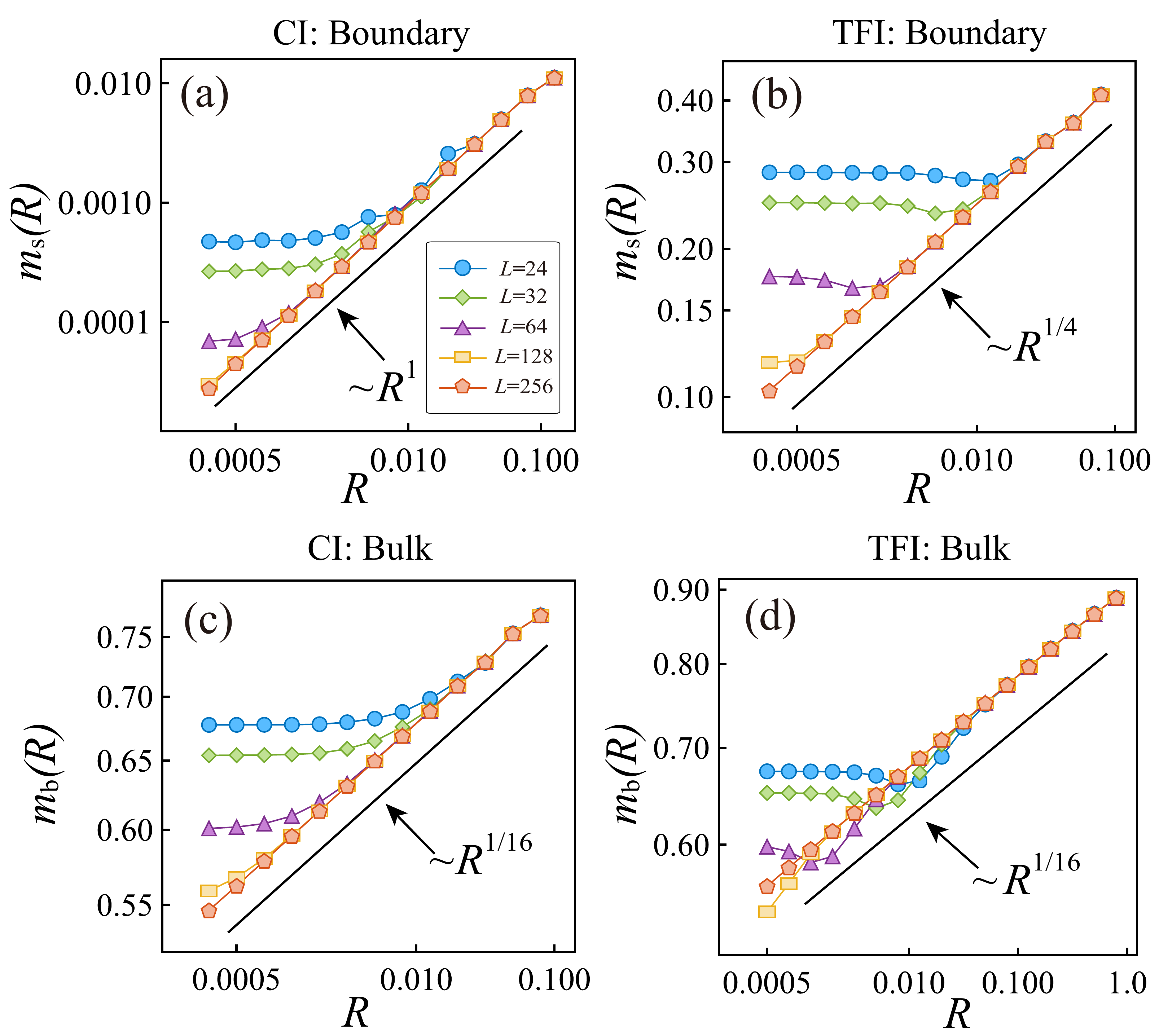}\\
	\caption{The boundary magnetization $m_s(R)$ at the end of the quench as a function of the quench rate $R$ for (a) the CI chain and (b) the TFI chain of  different system sizes. The dynamical scaling of $m_{s}(R)$ at QCPs exhibits power law scalings with distinctive exponents close to $1$ and $1/4$, respectively (solid lines). In contrast, the bulk magnetization $m_b$ in both (c) the CI chain and (d) the TFI chain exhibit the same power-law behavior, with an exponent close to 1/16 (solid lines).} \label{fig:2}
\end{figure}

According to the adiabatic-impulse approximation in KZ mechanism, when a system is linearly driven from initial ground state to the QCP, the system becomes frozen at the freeze-out time  $ \hat{t} \propto -R^{-z\nu/(1+z\nu)}$ with $z$ and $\nu$ being the dynamical and correlation-length exponents at the QCP. For the dimensionless distance to QCP ${\epsilon}\equiv R |t|$, the freeze-out time leads to $\hat{\epsilon}(\hat{t}) \propto R^{1/(1+z\nu)} $, and the physical quantities after the driven dynamics are controlled by $\hat{\epsilon}$. Consequently, the bulk magnetization $m_b$ and boundary magnetization $m_s$, which scale  as $m_{b, s}\propto \epsilon ^{\beta_{b, s}}$ in equilibrium,  are now expected to satisfy the dynamical scaling relation~\cite{Zhong2005PRB}:
\begin{equation}\label{eq:kz}
m_{b,s}(R) \propto \hat{\epsilon}^{\beta_{b,s}} \propto R^{\beta_{b,s}/(1+z\nu)}, 
\end{equation}
which was first obtained by using finite time scaling analysis~\cite{Zhong2005PRB,Gong2010NJP,Zhong2006PRE,Huang2014PRB,Huang2020PRR,Zeng2025NC,Huang2016PRB,Yin2016PRB,Shu2025NC,Yin2025arxiv,Zeng2025NC,Zeng2025PRB,shu2025universaldrivencriticaldynamics}. Here, $\beta_{b, s}$ are the critical exponents for bulk and boundary order-parameter, respectively. It indicates that the \emph{boundary} magnetization $m_s(R)$ should obey the standard KZ scaling by replacing the bulk exponent $\beta_b$ with the \emph{boundary} order-parameter exponent $\beta_s$ while the other exponents $\nu$ and $z$ remain governed by the universality class of the QCP.

In Fig.~\ref{fig:2}, we present numerical results for the scaling behaviors of both bulk and boundary magnetization in the TFI and CI chains following driven critical dynamics. In both models, the bulk magnetization exhibits universal scaling with an exponent of $1/16$  [Fig.~\ref{fig:2} (c) and (d)], consistent with the conventional KZ prediction (\ref{eq:kz}) when taking $\beta_b = 1/8$ and $z=\nu=1$ for the (1+1)-dimensional Ising universality class relevant here (see data collapse in SM Sec.~II ~\cite{sm}).

In contrast, the boundary dynamics reveal fundamentally different behaviors for the two models. As shown in Fig.~\ref{fig:2} (a) and (b), the boundary magnetization $m_s(R)$ in both models exhibit power-law scaling, but with distinct exponents. For the topologically trivial Ising QCP [Fig.~\ref{fig:2}(b)], $m_s(R)$ obeys the standard KZ scaling   (\ref{eq:kz}) controlled by the \emph{boundary} order-parameter exponent $\beta_s = 1/2$ \cite{Collura2009JOSM,McCoy1973TWo,diehl1997IJMP} and $z=\nu=1$, yielding an exponent $1/4$~\cite{shu2025universaldrivencriticaldynamics}.
Remarkably, at the topologically nontrivial $ \text{Ising}^{*} $ QCP [Fig.~\ref{fig:2} (a)], the boundary magnetization instead exhibits anomalous scaling $m_s(R) \propto R$ with an exponent $1$, which does not conform to the KZ prediction if taking into
account that $z=\nu=1$ and $\beta_s=1$ obtained from the equilibrium magnetization (see SM~\cite{sm}  Sec.~\Rmnum{2}).

To explain the anomalous scaling behavior of boundary magnetization, we propose a modified scaling relation: 
\begin{equation}\label{eq:rkz}
m_{b, s}(R) \propto R^{\Delta_{b, s}/r},
\end{equation}
in which $r=z+1/\nu$ represents the scaling dimension of the driving rate $R$ and $\Delta_{b, s}$ are the scaling dimensions of bulk and boundary order parameters, respectively. The scaling relation in Eq.~(\ref{eq:rkz}) is generic for (1+1)-dimensional QCPs described by conformal field theory (CFT) and can be derived from scaling arguments (see the detailed derivation and physical interpretation in the End Matter). Here, $ \Delta_b = \frac{1}{8} $ for both models, $ \Delta_s = \frac{1}{2} $ for TFI model, and $ \Delta_s = 2 $ for the CI model~\cite{Yu2022PRL}. These values account for all the scaling behaviors observed in Fig.~\ref{fig:2}.
For bulk magnetization,  the scaling dimension is connected to critical exponents $\Delta_b= \beta_b/\nu$, and Eq.~(\ref{eq:rkz}) reduces to the standard KZ scaling. For the boundary magnetization, the relation $\Delta_s=\beta_s/\nu$~\cite{D1997CON} continues to hold in the TFI model, but breaks down in the CI model due to the presence of robust topological edge states at criticality. Consequently, Eq.~\eqref{eq:rkz} not only reproduces the traditional KZ scaling behavior but also captures genuinely new behaviors beyond the KZ paradigm, thereby providing a more general description of driven critical dynamics. Essentially, these distinct boundary critical dynamics can be traced to the different topological properties of the QCPs, i.e., different conformal boundary conditions~\cite{Yu2022PRL,Parker2018PRB,Parker2019PRL}, which give rise to qualitatively different spontaneous boundary magnetizations.

\emph{Generalization to interacting quantum Potts chains.}---To further verify the generality of Eq.~\eqref{eq:rkz} beyond exactly solvable models, we consider a genuinely interacting system that cannot be reduced to a free-fermion description~\cite{Alica2016ARCMP}: a one-dimensional chain of qutrits represented by generalized Pauli operators
\begin{equation}
X = \begin{pmatrix}
0 & 1 & 0 \\
0 & 0 & 1 \\
1 & 0 & 0
\end{pmatrix}, \quad
Z = \begin{pmatrix}
1 & 0 & 0 \\
0 & \omega & 0 \\
0 & 0 & \omega^2
\end{pmatrix},
\end{equation}
where $\omega = e^{2\pi i/3}$ and $XZ = \omega ZX$.
We define two Hamiltonians on a bipartite chain with sublattices $A$ and $B$~\cite{Verresen2021PRX}:
\begin{equation}
\label{ph}
\begin{aligned}
H_{P1} &= -\sum_{n} Z_{B,n}^{\dagger} Z_{B,n+1}^{} - h_1 \sum_{n} (X_{A,n}^{} + X_{B,n}^{}) + \mathrm{H.c.} \, , \\
H_{P2} &= -\sum_{n} Z_{B,n}^{\dagger} Z_{B,n+1}^{} - h_2 \sum_{n} \Big( Z_{B,n-1}^{\dagger} X_{A,n}^{} Z_{B,n}^{} + \\
& \quad{}\,\, Z_{A,n}^{} X_{B,n}^{} Z_{A,n+1}^{\dagger}\Big) + \mathrm{H.c.} \, .
\end{aligned}
\end{equation}
Both Hamiltonians have a $\mathbb{Z}_3^A \times \mathbb{Z}_3^B$ symmetry generated by $P_A = \prod_n X_{A,n}$ and $P_B = \prod_n X_{B,n}$. By tuning $h_1$ ($h_2$), $H_{P1}$ ($H_{P2}$) undergoes a continuous quantum phase transition at $h_1=1$ ($h_2=1$) between a $\mathbb{Z}_3^B$ spontaneous symmetry breaking phase and a trivial product state (a $\mathbb{Z}_3^A \times \mathbb{Z}_3^B$ SPT phase), described by the (1+1)-dimensional Potts CFT with central charge $c=4/5$~\cite{D1997CON,Von1986JPA}. The associated local order parameter is given by the operator $(Z_{B,l}^{} + Z_{B,l}^{\dagger}) / 2$ for the $l$ -th site. Although the two transitions share the same bulk universality class, they are topologically distinct~\cite{Verresen2021PRX}: the QCP of $H_{P1}$ is topologically trivial with free boundary conditions, whereas the QCP of $H_{P2}$ is topologically nontrivial with spontaneously fixed boundary conditions, resulting in boundary magnetization even though the bulk remains gapless~\cite{Yu2022PRL,Parker2018PRB}.

\begin{figure}[t]
	\centering
\includegraphics[width=3.5in]{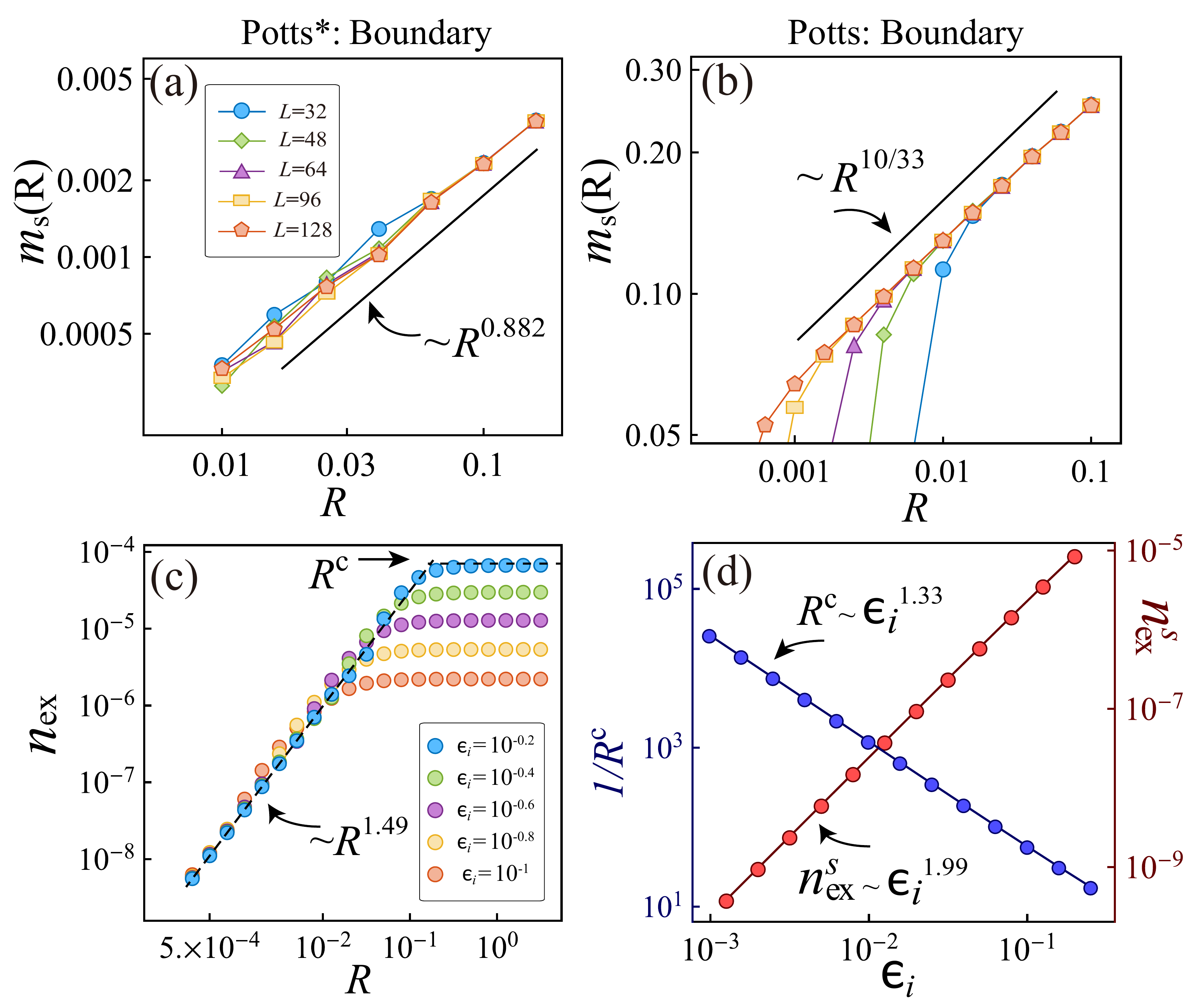}\\
	\caption{(a)-(b) Driven critical dynamics in the quantum Potts model.  The boundary magnetization $m_s(R)$ 
    at the end of the quench as a function of the quench rate $R$ for (a) the topologically nontrivial Potts (Potts*) QCP and (b) the trivial Potts (Potts) QCP. Different symbols correspond to different system sizes $L$. The dynamical scaling of $m_s(R)$ at topologically distinct QCPs exhibits power-law scalings with distinctive exponents close to $0.882$ and $10/33$, respectively (solid lines). (c)-(d) The anomalous dynamical scaling behaviors of the number of excitation $n_{\text{ex}}$ when the free-fermion system is driven to topologically nontrivial QCPs. (c) The dependence of  $n_{\text{ex}}$ on the quench rate $R$ for different values of $\epsilon_i$. In the slow-quench regime, $n_{\text{ex}}$ follows a power-law scaling with an exponent close to 1.5. In the fast-quench regime, $n_{\text{ex}}$ saturates to a value independent of the quench rate. (d) The saturation value  $ n^s_{\text{ex}}$ and the critical quench rate $R^c$ both exhibit power-law scaling with the dimensionless distance $\epsilon_i$, with exponents close to 2 and 4/3, respectively.
} \label{potts_dy}
\end{figure}

We performed tensor-network simulations~\cite{White1992PRL_DMRG,SCHOLLWOCK201196,Haegeman2011PRL,Haegeman2016PRB_TDVP,Yang2020PRB_TDVP} of the driven dynamics at these Potts QCPs, examining both bulk and boundary magnetizations using the same quench protocol as for the Ising chain. In both models, the bulk magnetization exhibits power-law scaling with exponent $2/33$ (see SM~\cite{sm} Sec.~II ), consistent with the KZ prediction using $\beta_b = 1/9$, $z=1$, and $\nu = 5/6$ ~\cite{D1997CON}. In contrast, the boundary magnetization shows qualitatively distinct behavior. At the trivial QCP [Fig.~\ref{potts_dy}~(b)], it follows standard KZ scaling with $\beta_s = 5/9$, yielding $m_s (R) \sim R^{10/33}$. Remarkably, at the topologically nontrivial QCP [Fig.~\ref{potts_dy}~(a)], the boundary magnetization exhibits anomalous scaling $m_s (R) \sim R^{0.882}$, deviating from the KZ prediction based on equilibrium exponents ($\beta_s \approx 0.952$, $z=1$, and $\nu=5/6$). Instead, given that $\Delta_b = 2/15$ for both models, $\Delta_s = 2/3$ for the trivial Potts QCP, and $\Delta_s = 1.94$ for the topologically nontrivial Potts QCP~\cite{D1997CON,CARDY1984514,CARDY1989581,Yu2022PRL}, all dynamical scaling behaviors are accurately captured by the modified scaling relation Eq.~(\ref{eq:rkz}). Numerical details and the determination of $\beta_{b,s}$ and $\Delta_{b,s}$ are presented in  SM~\cite{sm} Sec.~II.

\emph{Edge excitations at topological nontrivial QCPs.}---Gapless SPT physics can also emerge in the free-fermion systems, appearing at phase transition points between gapped phases distinguished by different nonzero topological indices~\cite{Verresen2018PRL,verresen2020topologyedgestatessurvive}. This motivates us to explore the dynamical scaling behaviors of topologically nontrivial QCPs in a broader class of models. These free-fermion analogues of gSPT physics are illustrated by the Jordan–Wigner dual of model~\eqref{E1}: 
\begin{equation}
\label{eq:fh}
\begin{aligned}
H_{\text{F}} & = g_0 H_0 - g_1 H_1 + g_2 H_2, \\
H_{\alpha} & = i \sum_{j} \gamma_{2j} \gamma_{2(j+\alpha)-1},
\end{aligned}
\end{equation}
where $\gamma_{2j-1}$ and $\gamma_{2j}$ denote two species of Majorana fermions on site $j$. The $H_{\text{F}}$ contains onsite ($g_0$), nearest-neighbor ($g_1$), and next-nearest-neighbor ($g_2$) hopping terms and exhibits several topologically distinct QCPs. The crucial difference in dynamical scaling among the free-fermion models is captured by the total number of edge excitations, $n_{\text{ex}} = \sum_n \sum_{m\in v} \big|\langle E_n(t_f) \mid \psi_m(t_f) \rangle\big|^2$\footnote{Although the definition of the number of excitations in the main text differs from that in the SM, where it is defined as the projection of the zero-energy edge state onto the excited (conduction-band) states of the final Hamiltonian, both definitions of $n_{\rm ex}$ yield identical scaling behavior, as demonstrated in SM~\cite{sm} Sec.~III.}, where the subscript $n$ labels the index of the energy spectrum, and $v$ denotes the  states belonging to the valence bands. 
Here, $|E_n(t_f)\rangle$ with $n = L, L+1$ denote the instantaneous eigenstates located near the center of the $2L$-level energy spectrum of the Hamiltonian at the final time $t_f$, and $|\psi_m(t_f)\rangle$ are obtained by time-evolving the initial valence-band eigenstates of the Hamiltonian at $t_i$ under the time-dependent Schrödinger equation(see SM~\cite{sm} Sec.~III for detailed explanations). We set $g_1 = 1$ and ramp $g_2$ linearly according to $g_2(t) = g_{2,i} + (g_{2,c} - g_{2,i}) R t, t \in [0, 1/R]$, driving the system to the topologically nontrivial QCP at $g_{2,c}$. Notably, a small onsite term $g_0 = 0.1$ is introduced to activate dynamical scaling by weakly coupling to the bulk degrees of freedom (see SM~\cite{sm} Sec.~\Rmnum{2} for a detailed discussion of this subtlety).

Fig.~\ref{potts_dy} (c) shows $n_{\text{ex}}$ as a function of the quench rate $R$ for various initial values of $\epsilon_i = (g_{2,i}-g_{2,c})/g_{2,c}$. Two distinct regimes are clearly visible. In the slow quench limit, the edge-excitation number exhibits a anomalous power-law scaling, $n_{\text{ex}} \propto R^{1.49}$. 
In the fast-quench regime, $n_{\text{ex}}$ approaches a plateau whose saturated value $n_{\text{ex}}^{ s}$  is independent of $R$. We further find that this saturated value exhibits an additional power-law scaling with the initial parameter $\epsilon_i < 1$ [Fig.~\ref{potts_dy} (d)], a feature absent at topologically trivial QCPs. Moreover, the critical quench rate $R^{c}$—defined as the crossover point between the two regimes in Fig.~\ref{potts_dy} (c)—scales as $R^{c} \propto \epsilon_i^{1.33}$, as shown in Fig.~\ref{potts_dy} (d). 
For comparison, we also investigate driven dynamics at a topologically trivial QCP by applying a similar linear ramp in $g_0(t)$ while fixing $g_1 = 1$ and $g_2 = 0$. The resulting edge-excitation number, shown in the End Matter, saturates at $n^s_{\text{ex}} \sim 1.0$ and exhibits no power-law scaling, since the edge modes delocalize and merge into the gapless bulk at the trivial QCP~\cite{Bermudez_2010}. Taken together, these results indicate that robust topological edge states at criticality qualitatively modify the dynamics  and give rise to anomalous scaling behaviors at topological quantum criticality.

We remark that the mechanisms underlying anomalous dynamical scaling in interacting spin chains and free-fermion models are fundamentally different. In interacting spin chains, it originates from emergent spontaneous boundary symmetry breaking at topologically nontrivial QCPs, giving rise to unconventional surface critical behavior characterized by boundary scaling dimensions. In contrast, spontaneous symmetry breaking is absent in free-fermion systems, and the dynamical scaling of the number of excitations $n_{\rm ex}$ is unrelated to any boundary scaling dimension. Physically, at a topologically nontrivial QCP, the edge states remain localized near the boundary with a finite localization length and do not hybridize with the gapless bulk modes. As a result, excitations from bulk states into these localized edge modes are suppressed compared to the topologically trivial case, where the edge states delocalize into the bulk. Therefore, the generation of excitations exhibit totally different scaling behaviors for the topologically trivial QCPs ($n_{\text{ex}} \sim R^{0}$) and nontrivial QCPs ($n_{\text{ex}} \sim R^{1.49}$).

\emph{Concluding remarks.}---To conclude, we have uncovered a previously unrecognized mechanism through which nontrivial topology at QCPs fundamentally modifies driven critical dynamics and gives rise to anomalous scaling behaviors. By examining the driven dynamics of bulk and boundary magnetization at topologically distinct QCPs in two families of quantum spin chains—one exactly solvable and the other not—we find that while the local bulk dynamics remain indistinguishable and follow standard KZ scaling, the boundary dynamics exhibit a  universal scaling behavior unique to topological criticality and inaccessible within the conventional KZ framework. We further investigate driven dynamics at topologically distinct QCPs in free-fermion models and observe similar anomalous scaling in defect production at topologically nontrivial QCPs. The topology-induced defect production remains stable against symmetry-preserving disorder (see End Matter) and is inherently tied to the presence of topological edge modes at criticality, providing a novel and universal mechanism for realizing anomalous dynamical scaling.

Looking ahead, it would be highly intriguing to explore the dynamical properties of topologically nontrivial QCPs in generic interacting systems, particularly in higher dimensions, which are expected to exhibit far richer topological phenomena than one-dimensional counterparts~\cite{Wen2017RMP}. Furthermore, the topology-induced anomalous dynamical scaling identified in this work could serve as a natural and practical probe of topological features at criticality, particularly in settings where preparing the true critical ground state remains challenging on modern quantum platforms.

\textit{Acknowledgement}: 
This work was supported by the National Natural Science Foundation of China (Grants No.12275075, No.12405034), the National Key Research and Development Program of Ministry of Science and Technology (No.2021YFA1200700), China Postdoctoral Science Foundation (Certificate Number: 2024M752760), and the start-up grant from Eastern Institute of Technology, Ningbo. Tensor-network simulations were carried out with the ITENSOR \verb|C++| package~\cite{itensor}.

\bibliographystyle{apsrev4-2}
\let\oldaddcontentsline\addcontentsline
\renewcommand{\addcontentsline}[3]{}
\bibliography{ref.bib}

\appendix
\section{\large{End Matter}}
\twocolumngrid

\textit{Detailed derivation of Eq.~(4) based on scaling arguments.}---For driven dynamics near a critical point in one-dimensional quantum spin chains described by conformal field theory (CFT), starting from an ordered initial state, the order parameter $O$ (e.g., the magnetization in the quantum spin chain) with scaling dimension $\Delta_O$ obeys the scaling form~\cite{D1997CON,LeePRL2014,PelissettoPRB2018}:
\begin{equation}
O(\epsilon, L, R) = b^{-\Delta_O} F(\epsilon b^{1/\nu}, L b^{-1}, R b^r),
\label{eq:rpres1}
\end{equation}
where $\epsilon$ measures the distance from the critical point, $L$ is the system size, $R$ is the quench rate, and $b$ is the rescaling factor. Here, $\nu$ is the correlation-length exponent, $r = z + 1/\nu$ denotes the scaling dimension of $R$ ($z$ is the dynamical critical exponent), and $F$ is a non-singular scaling function.

The scaling combinations $\epsilon b^{1/\nu}$ and $R b^r$ follow naturally from the linear quench protocol $\epsilon = R t$. Under rescaling, time transforms as $t’ = t / b^z$, consistent with the relaxation time $\tau \sim \xi^z$, while $\epsilon$ scales as $\epsilon’ = \epsilon b^{1/\nu}$ due to $\xi \sim |\epsilon|^{-\nu}$. Substituting $\epsilon = R t$ then yields $R’ = R b^{z + 1/\nu}$, which defines $r = z + 1/\nu$.

At the critical point ($\epsilon = 0$) in the thermodynamic limit ($L \to \infty$), choosing $R b^r = 1$ leads to
\begin{equation}
O(R) \propto R^{\Delta_O / r}.
\label{eq:rpres2a}
\end{equation}
which corresponds to Eq.~(4) in the main text. This relation follows directly from scaling arguments and is generally valid for (1+1)-dimensional quantum critical points (QCPs) described by CFT.

{\it Comparison with the conventional Kibble-Zurek scaling.}- For comparison, we also present the conventional Kibble–Zurek (KZ) scaling prediction: 
 \begin{equation}\label{eq:rpres4a}
 O(R) \propto R^{\beta_O / (1 + z \nu)}.
 \end{equation} 
where $\beta_O$ is the order-parameter critical exponent. We would like to emphasize the relation and distinction between the scaling forms Eqs. (\ref{eq:rpres2a}) and (\ref{eq:rpres4a}). At first glance, the two forms appear equivalent through the relation $\Delta_O = \beta_O / \nu$~\cite{D1997CON}
. This relation is often regarded as universal and is commonly assumed to hold for the dynamical critical scaling of bulk order parameters~\cite{Gong2010NJP}, and has recently been generalized to boundary order parameters~\cite{shu2025universaldrivencriticaldynamics}. However, this equivalence breaks down when nontrivial topological edge modes at the QCP are taken into account.

Specifically, for the driven critical dynamics of bulk order parameters, the relation $\Delta_{O}^{\mathrm{b}} = \beta_{O}^{\mathrm{b}} / \nu$ generally holds, and Eq.~(\ref{eq:rpres2a}) reduces to the conventional KZ scaling Eq.~(\ref{eq:rpres4a}), as illustrated in Fig. 2 (c) and (d) of the main text for the critical cluster Ising chain and the critical transverse-field Ising chain, respectively. For the driven critical dynamics of boundary order parameters, if the QCP is topologically trivial (i.e., it does not host topological edge states), such as in the transverse-field Ising chain, the same relation remains valid, with $\Delta_O$ and $\beta_O$ replaced by the corresponding boundary quantities $\Delta_{O}^{\mathrm{s}}$ and $\beta_{O}^{\mathrm{s}}$. This corresponds to the KZ scaling for boundary order parameters, which has been recently investigated~\cite{shu2025universaldrivencriticaldynamics} and is also confirmed in Fig. 2 (b) of the main text.

In contrast, when the QCP hosts topological edge states and is therefore topologically nontrivial—this being the main focus of our work—the expected relation $\Delta_{O}^{\mathrm{s}} = \beta_{O}^{\mathrm{s}} / \nu$ breaks down. In this case, only Eq.~(\ref{eq:rpres2a}) correctly captures the dynamical scaling of the boundary order parameter, indicating that Eq. (\ref{eq:rpres2a}) is more general and extends beyond the conventional KZ scaling framework.

From a physical perspective, the origin of this anomalous scaling behavior can be understood as follows. For the driven critical dynamics of bulk order parameters, irrespective of the topological nature of the critical point, 
the critical fluctuations are fully governed by the divergence of the bulk correlation length $\xi \propto |\epsilon|^{-\nu}$. Consequently, the relation 
$\Delta_{O}^{\mathrm{b}} = \beta_{O}^{\mathrm{b}} / \nu$ always holds. For the driven critical dynamics of boundary order parameters, if the QCP is topologically trivial, the boundary fluctuations remain dominated by the divergent bulk critical fluctuations through bulk-boundary coupling, and the analogous relation $\Delta_{O}^{\mathrm{s}} = \beta_{O}^{\mathrm{s}} / \nu$ continues to apply, with $\Delta_{O}^{\mathrm{b}}$ and $\beta_{O}^{\mathrm{b}}$ replaced by their boundary counterparts $\Delta_{O}^{\mathrm{s}}$ and $\beta_{O}^{\mathrm{s}}$.

However, at a topologically nontrivial QCP, the presence of edge states implies that boundary fluctuations are no longer governed by bulk critical fluctuations. Consequently, the bulk correlation-length exponent $\nu$ is no longer the appropriate quantity entering the relation between scaling dimension and order parameter exponent. We emphasize that, because the quench is applied uniformly to all sites of the spin chain, the scaling dimension $r = z + 1/\nu$ associated with the quench rate $R$ remains position independent, and the bulk exponent $\nu$ still enters $r$, regardless of the topological nature of the QCP. In this case, the boundary KZ scaling form $O_s(R) \propto R^{\beta_O^s / (1 + z \nu)}$ breaks down, and the correct scaling behavior is instead given by $O_s(R) \propto R^{\Delta_O^s / r}$, as verified in Fig.~2 (a) of the main text. Therefore, the scaling relation in Eq.~(\ref{eq:rpres2a}) not only reproduces the standard KZ scaling but also captures genuinely new behavior beyond the KZ paradigm, providing a more general description of driven critical dynamics. 

\begin{figure}[t]
	\centering
\includegraphics[width=3.5in]{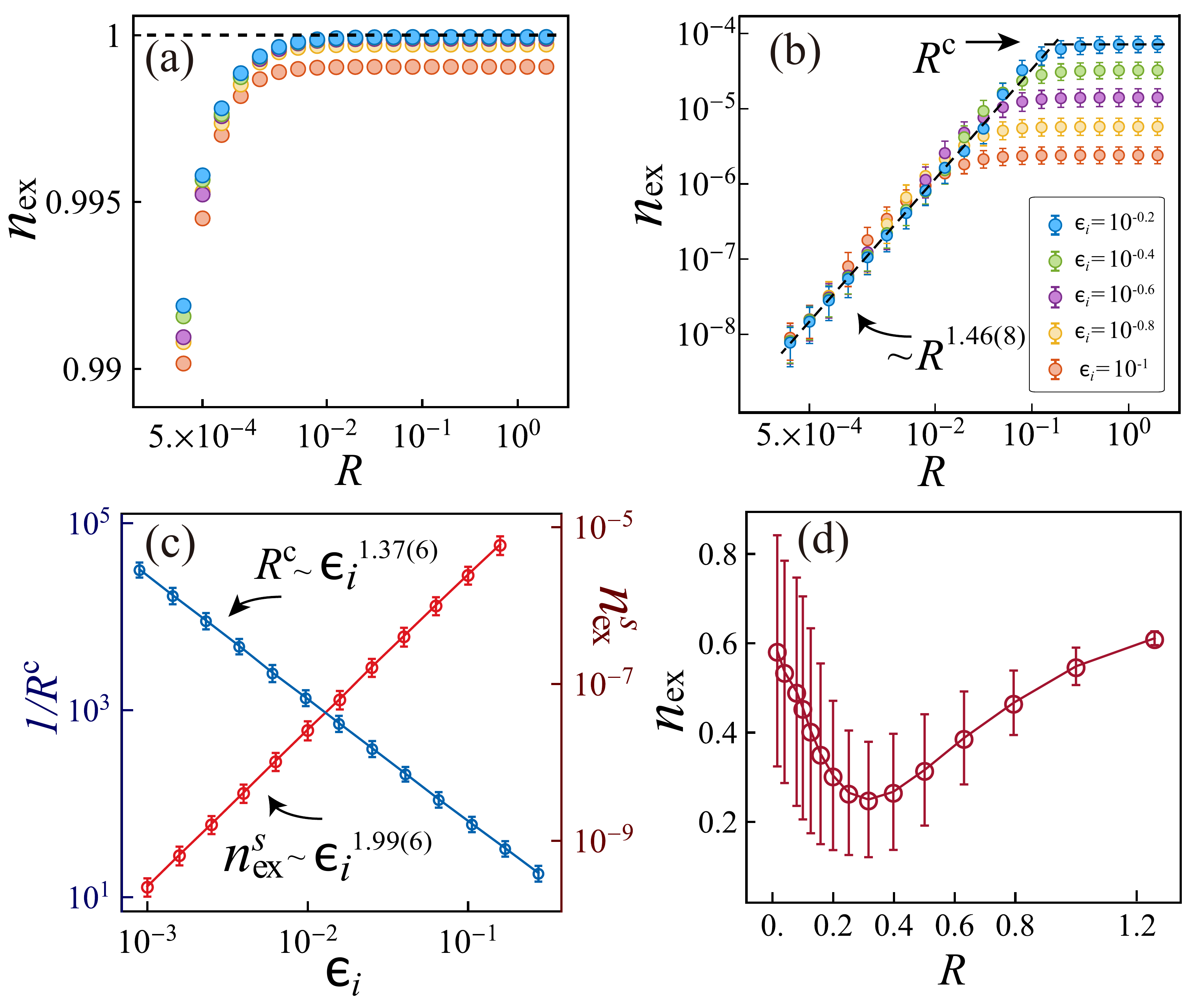}\\
\caption{  (a) The excitations $n_{\text{ex}}$ generated by quenching to the topologically trivial  QCP as a function of the quench rate $R$ for different values of $\epsilon_i$ in the clean system. 
 (b)-(c) The anomalous dynamical scaling behaviors of the excitation number $n_{\text{ex}}$ when the system is driven to topologically nontrivial QCPs in the presence of symmetry-preserving disorder. 
 (b) In the slow-quench regime, $n_{\text{ex}}$ follows a power-law scaling with an exponent close to 1.5. In the fast-quench regime, $n_{\text{ex}}$ saturates to a value independent of the quench rate. 
 (c) The saturation value $n_{\text{ex}}^s$ and the critical quench rate $R^c$ both exhibit power-law scaling with the dimensionless distance $\epsilon_i$, with exponents close to 2 and 4/3, respectively. The anomalous dynamical power-law scaling remains robust and exhibits the same  exponents as the clean system (Fig.~\ref{potts_dy}(c) and (d)) within error bars. 
 (d) Across the critical point of the Creutz ladder model, the topology-induced anomalous power-law behavior of defect production reported in Ref.~\cite{Bermudez2009PRL} is destroyed by symmetry-preserving disorder. Error bars represent standard deviations. The size of the clean system is $L=1200$, while the disordered system has $L=400$. Averages were performed over 600 independent disorder realizations.
} \label{nex_dis}
\end{figure}

\textit{Further discussion of the anomalous scaling of edge excitations.}---To examine the stability of topology-induced anomalous dynamics in the main text, we consider the 1D critical free-fermion model [Eq.~\eqref{eq:fh} in the main text] with symmetry-preserving disorder in $g_{0/1}^{\text{dis}}$, drawn independently at each site from a uniform distribution $g_{0/1}^{\text{dis}} \in g_{0/1}[1-\delta,1+\delta]$ with disorder strength $\delta = 0.1$. We then linearly ramp the parameter $g_2$ toward the disordered topologically nontrivial QCP. The resulting edge-excitation number $n_{\text{ex}}$ as a function of the quench rate $R$ is shown in Fig.~\ref{nex_dis} (b). The numerical results clearly demonstrate that, in the slow-quench regime, $n_{\text{ex}}$ continues to exhibit the anomalous power-law scaling $n_{\text{ex}} \propto R^{1.46(8)}$. In the fast quench regime, the same universal behaviors observed in the clean case persist (Fig.~\ref{nex_dis} (c)): the edge excitation number saturates to a value that scales as $n^s_{\text{ex}} \propto \epsilon_i^{1.99(6)}$, and the critical quench rate follows $R^c \propto \epsilon_i^{1.37(6)}$. Therefore, the dynamical scaling exponents at disordered QCPs are consistent with those of the clean case within error bars, indicating that the anomalous dynamical scaling is robust against symmetry-preserving disorder. We further verify that, under sufficiently strong disorder, the critical system flows to the infinite-randomness fixed point characterized by the effective central charge $c_{\text{eff}} = \ln 2$, i.e., the random-singlet fixed point where disorder flows to infinity~\cite{Fisher1994PRB,Fisher1995PRB,Refael2004PRL} (see Sec.~\Rmnum{2} of the SM~\cite{sm}). Remarkably, despite the flow to infinite randomness, the anomalous dynamical scaling remains robust as long as the topological edge modes remain stable at the QCP.

We further emphasize a key distinction between our work and the topology-induced anomalous defect production discussed previously in the literature~\cite{Bermudez2009PRL}. In those studies, the reported anomalous dynamics arise from special features of the Creutz model~\cite{Creutz1999PRL} (see Sec.~\Rmnum{2} of the SM~\cite{sm} for a brief review) and are unstable against symmetry-preserving disorder, as shown in Fig.~\ref{nex_dis} (d). Indeed, the critical points separating the topologically distinct phases in the Creutz model do not host stable (i.e., non fine-tuned) topological edge modes and are therefore unrelated to the intrinsic topological properties of the QCP. This stands in sharp contrast to our case: the anomalous dynamical behavior we identify is inherently tied to the presence of robust topological edge modes at the QCP itself.

To further substantiate that these anomalous dynamics originate from the topological edge modes at criticality, we investigate in Sec.~\Rmnum{2} of the SM ~\cite{sm} the driven dynamics at topologically nontrivial QCPs for various strengths of the bulk coupling $g_0$. The results unambiguously show that when $g_0$ is smaller than a threshold value $g_{0}^{*}$, the topological edge modes persist and coexist with the gapless bulk, leading to \emph{universal} anomalous power-law scaling of $n_{\text{ex}}$ with $R$, characterized by the same exponent $\sim 1.5$. In contrast, when $g_0$ exceeds $g_{0}^{*}$, the edge modes  delocalize and merge into the gapless bulk, and the dynamical scaling reverts to the usual behavior expected at trivial QCPs, $n_{\text{ex}} \sim 1.0$.

The scaling exponents associated with the anomalous dynamics are \emph{universal} and independent of the number of edge modes at topologically nontrivial QCPs in free-fermion models with general $\alpha>2$ (see Sec.~\Rmnum{4} of the SM~\cite{sm}). More intriguingly, anomalous dynamical scaling also emerges at two-dimensional topological Chern criticality (see Sec.~\Rmnum{5} of the SM~\cite{sm}). These results demonstrate that the presence of robust topological edge states at criticality constitutes a new underlying mechanism that fundamentally modifies the dynamics and gives rise to anomalous scaling laws. We further remark that although the main text focuses on driven dynamics toward the critical points, anomalous dynamical scaling can also arise in protocols that drive the system \emph{across} topologically nontrivial QCPs (see Sec.~\Rmnum{2} of the SM~\cite{sm} for details), making this phenomenon more readily accessible in modern experimental platforms.

\let\addcontentsline\oldaddcontentsline
\onecolumngrid

\clearpage
\newpage

\widetext

\begin{center}
\textbf{\large Supplemental Material for ``Anomalous Dynamical Scaling at Topological Quantum Criticality''}
\end{center}

\maketitle

\renewcommand{\thefigure}{S\arabic{figure}}
\setcounter{figure}{0}
\renewcommand{\theequation}{S\arabic{equation}}
\setcounter{equation}{0}
\renewcommand{\thesection}{\Roman{section}}
\setcounter{section}{0}
\setcounter{secnumdepth}{4}

\addtocontents{toc}{\protect\setcounter{tocdepth}{0}}
{
\tableofcontents
}

\section{A brief review of Gaussian state method for calculating dynamical scaling under open boundary conditions}
\label{sec:1}

The transverse field spin-$\frac{1}{2}$ chain with cluster interaction can be mapped to non-interacting Majorana fermions using the Jordan-Wigner transformation \cite{jones2019asymptotic}, $\sigma_l^z = \prod_{k=1}^{l-1}(i\gamma_{2k-1}\gamma_{2k})\gamma_{2l-1}$ and $\sigma_l^x = i \gamma_{2l-1}\gamma_{2l},$
in which the Majorana fermion operators satisfy the anticommutation relation $\{\gamma_k, \gamma_l\} = 2\delta_{kl}$. In this Majorana representation, the Hamiltonian from Eq.~(1) in the main text takes the form:
\begin{equation}
\label{eq:SE1}
H_{\text{fermion}} = -J\sum_{l=1}^{L-1} i \gamma_{2l}\gamma_{2l+1} - h\sum_{l=1}^{L} i \gamma_{2l-1}\gamma_{2l} - J_3\sum_{l=1}^{L-2} i \gamma_{2l}\gamma_{2l+3}.
\end{equation}
To diagonalize the Hamiltonian (\ref{eq:SE1}), we introduce the complex fermion operators $c_l$ and $c_l^{\dagger}$: $\gamma_{2l-1} = c_l^{\dagger} + c_l$ and $\gamma_{2l} = i(c_l^{\dagger} - c_l).$

Consequently, the Hamiltonian can be expressed as:
\begin{equation}
\label{eq:SE2}
H_{\text{fermion}} = \sum_{i,j} (A_{ij} c_i^{\dagger} c_j + A_{ij}^{\ast} c_j^{\dagger} c_i) + \sum_{i,j} (B_{ij} c_i^{\dagger} c_j^{\dagger} + B_{ij}^{\ast} c_j c_i).
\end{equation}
Here, $A = A^{\dagger}$ is Hermitian, and $B = -B^{\dagger}$ is antisymmetric, with the following definitions:
\begin{equation}
\begin{cases}
A_{ij} = -h \delta_{i,j} + \frac{J}{2}\delta_{i,j+1} + \frac{J}{2}\delta_{i+1,j} + \frac{J_3}{2}\delta_{i,j+2} + \frac{J_3}{2}\delta_{i+2,j};\\
\\
B_{ij} = \frac{J}{2}\delta_{i+1,j} - \frac{J}{2}\delta_{i,j+1} + \frac{J_3}{2}\delta_{i+2,j} - \frac{J_3}{2}\delta_{i,j+2}.
\end{cases}
\end{equation}

We then perform a canonical transformation, defined as:
\begin{equation}
\label{Eq:SE4}
\begin{cases}
\eta_{\mu} = \sum_{j=1}^{L} (U_{j\mu}^{\ast} c_j + V_{j\mu}^{\ast} c_j^{\dagger}); \\
\\
\eta_{\mu}^{\dagger} = \sum_{j=1}^{L} (V_{j\mu} c_j + U_{j\mu} c_j^{\dagger}).
\end{cases}
\end{equation}
In this context, $U_{j\mu}$ and $V_{j\mu}$ satisfy the Bogoliubov-de Gennes equations:
\begin{equation}
\begin{cases}
\sum_k (A_{jk} U_{k\mu} + B_{jk} V_{k\mu}) = \Lambda_{\mu} U_{j\mu}; \\
\\
\sum_k (B^{\ast}_{jk} U_{k\mu} + A^{\ast}_{jk} V_{k\mu}) = -\Lambda_{\mu} V_{j\mu}.
\end{cases}
\end{equation}
The Hamiltonian is thus diagonalized as:
\begin{equation}
\label{SE2}
H_{\text{F}} = \sum_{\mu} \Lambda_{\mu} \left(\eta_{\mu}^{\dagger} \eta_{\mu} - \frac{1}{2}\right),
\end{equation}
where the eigenenergies satisfy $\Lambda_{\mu} \geq 0$, and the ground state is the state annihilated by all $\eta_\mu$, denoted as $|GS\rangle$: $\eta_{\mu}|GS\rangle = 0, \forall \mu$.

\emph{\textbf{Local magnetization.}}---
We begin by discussing the calculation of local magnetization in equilibrium. The local magnetization referenced in the main text is longitudinal, which, in any eigenstate of Eq.~(\ref{eq:SE1}), is zero according to Wick's theorem. This is due to the operators $c_i$ and $c_i^{\dagger}$, which can be linearly mapped onto $\eta_i$ and $\eta_i^{\dagger}$, appearing an odd number of times in $\sigma_i^z$. However, one may consider introducing an infinitesimally small symmetry-breaking field along the spin interactions. Such a field will mix the two lowest eigenstates such that the resulting superposition can be expressed as:
\begin{equation}
|\psi\rangle = \frac{|GS\rangle + \eta_1^{\dagger}|GS\rangle}{\sqrt{2}}.
\end{equation}
This leads to the following expression for the local longitudinal magnetization \cite{Igloi1998PRB,Platini2007JPA,Proniak2019PS}:
\begin{equation}
\label{SE2}
m_l = |\langle GS|\eta_1\sigma_l^z|GS\rangle|.
\end{equation}
The operator $\eta_1\sigma^z_l$ can be conveniently written  as 
\begin{equation}
\label{eq:SE9}
\eta_1\sigma^z_l = \eta_1 A_1 B_1 A_2 B_2 \cdots A_{l-1} B_{l-1} A_l, \quad A_i = c_i + c_i^{\dagger}, \quad B_i = c_i - c_i^{\dagger}.
\end{equation}
Using Wick's theorem again, one can show that 
\begin{equation}
\langle GS | \eta_1 \sigma_{l}^z | GS\rangle = \text{Pf}(C),
\end{equation}
where the matrix $C$ is given by
\begin{equation}
\label{eq:Pf}
C =
\begin{pmatrix}
0 & \langle \eta_1 A_1 \rangle & \langle \eta_1 B_1 \rangle & \langle \eta_1 A_2 \rangle & \cdots & \langle \eta_1 A_l \rangle \\
& 0 & \langle A_1 B_1 \rangle & \langle A_1 A_2 \rangle & \cdots & \langle A_1 A_l\rangle \\
& & 0 & \langle B_1 A_2 \rangle & \cdots & \langle B_1 A_l \rangle \\
& & & \ddots & \vdots & \vdots \\
& & & & 0 & \langle B_{l-1} A_l \rangle \\
& & & & & 0
\end{pmatrix},
\end{equation}
in which Pf denotes the Pfaffian. The lower triangle of the matrix $C$ can be obtained from the relation $C = -C^{T}$, and the expectation values are evaluated in the ground state $|GS\rangle$. By combining the inverse transformation of Eq.~(\ref{Eq:SE4}) with (\ref{eq:SE9}), we can compute the corresponding matrix elements of Eq.~(\ref{eq:Pf}). Pfaffians of skew-symmetric matrices can be efficiently computed using the Householder transformation~\cite{Wimmer2012ACM}.

It is noteworthy that, for the CI chain, the presence of spontaneous edge magnetization leads to a two-fold degeneracy of its ground state, resulting in the magnetization at the first site remaining fixed at unity. Consequently, we focus on the local magnetization at the second site. Conversely, for the TFI chain, we continue to calculate the magnetization at the first site. Additionally, because of the edge magnetization in the CI chain, \( |\langle GS|\eta_1\sigma_2^z|GS\rangle| \) at the QCP tends toward a non-zero value even in the thermodynamic limit $L \to \infty$. Thus, we focus on the deviation of the boundary magnetization from its thermodynamic value in the CI chain, which  is given by $m_s = |\langle GS|\eta_1\sigma_2^z|GS\rangle| - |\langle GS|\eta_1\sigma_2^z|GS\rangle|_{L \to \infty}.$
For the bulk of the CI chain, as well as for  both the boundary and bulk of the TFI chain, $|\langle GS|\eta_1\sigma_l^z|GS\rangle|_{L \to \infty}$ at the QCP approach zero~\cite{Igloi2011PRL}. In the numerical calculation of $|\langle GS|\eta_1\sigma_2^z|GS\rangle|_{L \to \infty} $, we set $L = 9000$, which is considerably larger than the system sizes  studied in the following.


Next, we present the calculation of the time-dependent local magnetization. For convenience, we work in the Heisenberg picture. The local magnetization is given by: 
\begin{equation}
m_l(t) = |\langle GS|\eta_{1,H}(t)\sigma_{l,H}^z(t)|GS\rangle|,
\end{equation}
where $|GS\rangle$ is the initial ground state at $t=t_0$. The matrix element in this equation can be computed similarly to Eq.~(\ref{eq:Pf}), except that the operators $A_{j,H}(t) = c_{j,H}(t) + c^{\dagger}_{j,H}(t)$ and $B_{j,H}(t) = c_{j,H}(t) - c_{j,H}^{\dagger}(t)$ are now time-dependent. Therefore, we need to know their time evolution, described by the Heisenberg equation for the $c$ operators:
\begin{equation}
i\hbar \frac{d}{dt} c_{j,H}(t) = \left[ c_{j,H}, H_{j,H}(t) \right].
\end{equation}
This can be rewritten as
\begin{equation}\label{eq:hse2}
i\hbar \frac{d}{dt} c_{j,H}(t) = 2 \sum_{j'=1}^{L} \left[ A_{jj'}(t) \, c_{j',H}(t) + B_{jj'}(t) \, c_{j',H}^\dagger(t) \right].
\end{equation}
The time-dependent Bogoliubov method assumes that the Heisenberg operators associated with the Bogoliubov fermions are time-independent, coinciding with the original $\eta_{\mu}$ at time $t=t_0$, i.e., $\eta_{\mu} = \eta_{\mu,H} = \eta_{\mu}(t=t_0)$~\cite{Dziarmaga2005PRL,Glen2024SP}. Thus, the time-dependent Bogoliubov transformation yields:  
\begin{equation}\label{eq:tbg}
c_{j,H}(t) = \sum_{\mu=1}^{L} \left( U_{j\mu}(t)  \eta_\mu + V_{j\mu}^{\ast}(t)  \eta_\mu^\dagger \right),
\end{equation} 
with the initial conditions $U_{j\mu}(t_0)=U_{j\mu}$ and $V_{j\mu}(t_0)=V_{j\mu}$.

Substituting Eq.~(\ref{eq:tbg}) into the Heisenberg equation (\ref{eq:hse2}) yields a set of linearly coupled ordinary differential equations:
\begin{equation}
\begin{cases}
i\hbar \frac{d}{dt} U_{j\mu}(t) = 2 \sum_{j'=1}^{L} \left( A_{jj'}(t) U_{j'\mu}(t) + B_{jj'}(t) V_{j'\mu}(t) \right), \\ 
i\hbar \frac{d}{dt} V_{j\mu}(t) = -2 \sum_{j'=1}^{L} \left( B_{jj'}^{*}(t) U_{j'\mu}(t) + A_{jj'}^{*}(t) V_{j'\mu}(t) \right).
\end{cases}
\end{equation}
The functions $U_{j\mu}(t)$ and $V_{j\mu}(t)$ can be utilized to compute all correlation functions necessary for obtaining the time-dependent local magnetization $m_l(t)$ from the time-dependent version of Eq.~(\ref{eq:Pf}). Furthermore, we are particularly interested in the edge magnetization when the system reaches the QCP. In the limit of $R \rightarrow 0$, $m_l(R)$ approaches its equilibrium value $m_l^{eq}$  at criticality. As mentioned earlier, for the CI chain, due to the presence of edge magnetization, $m_s^{eq}$ at QCPs tends toward a non-zero value even in the thermodynamic limit. Hence, for the CI chain, the edge magnetization is expressed as $ m_s(t) = |\langle GS | \eta_{1,H} \sigma_{s,H}^z(t) | GS \rangle| - m_s^{eq}$. In the thermodynamic limit, the equilibrium bulk magnetization of the CI chain, along with both the edge and bulk magnetizations $m_{s,b}^{eq}$ of the TFI chain, all approach zero at the QCP.

\emph{\textbf{Entanglement entropy.}}---In this subsection, we review the computation of entanglement entropy for free fermion systems. For fermionic biquadratic (static) Hamiltonians, the density matrix can be derived from correlation functions~\cite{Vidal2003PRL,Ingo2003JPA,Canovi2014PRB}. As previously mentioned, we have introduced Majorana fermions $\gamma_{2l-1}$ and $\gamma_{2l}$. The correlation matrix of the Majorana fermions is given by
\begin{equation}
\label{SE2}
\langle \gamma_m \gamma_n \rangle = \delta_{mn} + i \Gamma_{mn},
\end{equation}
where \( m, n = 1, \ldots, 2l \). The matrix \( \Gamma_{mn} \) is antisymmetric, and its eigenvalues are purely imaginary, denoted as \( \pm i \nu_r \) for \( r = 1, \ldots, l \). It can be shown that this matrix describes a set of uncorrelated  fermions \( \{ d_m \} \) that satisfy
\begin{equation}
\label{SE3}
\langle d_m d_n \rangle = 0, \quad \quad \quad \langle d_m^{\dagger} d_n \rangle = \delta_{mn} \frac{1 + \nu_n}{2}.
\end{equation}
Each of the \( l \) blocks is then in the state $\rho_j = p_j d_j^{\dagger} |0\rangle \langle 0| d_j + (1 - p_j) |0\rangle \langle 0|,$
with \( p_j = \frac{1 + \nu_j}{2} \), such that the entropy is the sum of the single-particle entropies. Consequently, for the reduced \( l \)-site system, the entanglement entropy is given by
\begin{equation}
S(l) = \sum_{j=1}^{l} H_2 \left( \frac{1 + \nu_j}{2} \right),
\end{equation}
where \( H_2(x) = -x \log_2 x - (1-x) \log_2(1-x) \). The eigenvalues \( \lambda_j, \quad j = 1, \ldots, 2^{l} \) of the reduced density matrix can, in principle, be determined by taking appropriately chosen products of either $p_j$  or $(1 - p_j)$, with $j = 1, \ldots, l$~\cite{Calabrese2008PRA}.

\section{Additional numerical data for the anomalous dynamics at topologically distinct QCPs in spin and free-fermion models.}
\label{sec:2}

\begin{figure}[t]
	\centering
\includegraphics[width=7in]{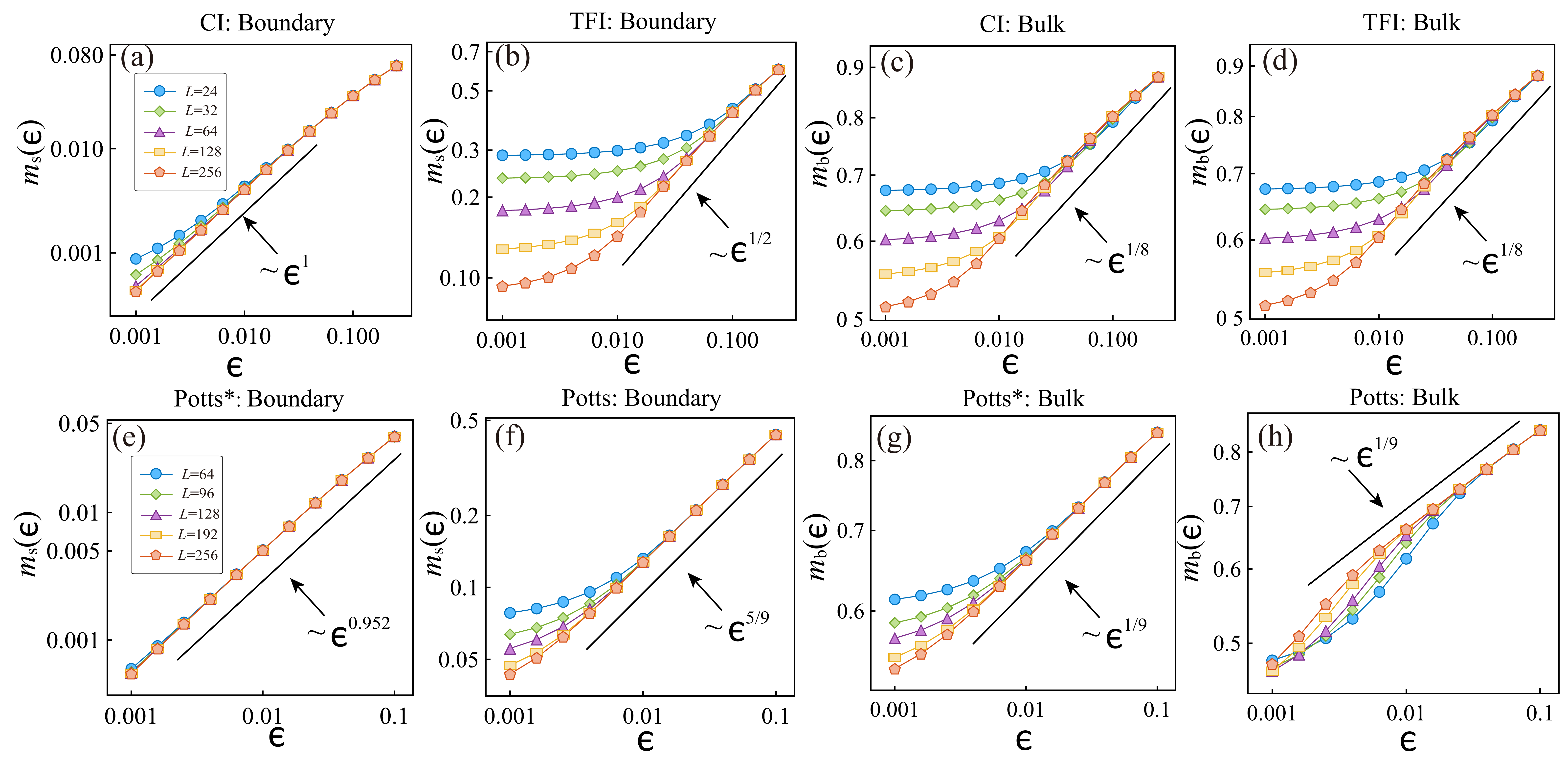}\\
	\caption{In equilibrium, the local magnetization $m_{b,s}$  as a function of the dimensionless distance $\epsilon$ to the QCP from topologically distinct QCPs. The black solid lines provide the guidance for an eye for the expected scaling (slope) on the log-log plot. (a)-(d)  Numerical results for topologically distinct Ising QCPs. The corresponding slopes approach the values of (a) $1$, (b) $1/2$, (c) $1/8$, and (d) $1/8$, respectively. (e)-(h) Same as (a)-(d) but for topologically distinct Potts QCPs. The corresponding slopes approach the values of (e) 0.952, (f) 5/9, (g) 1/9, and (h) 1/9, respectively. In (f) and (h), we have added $-(Z_{B,L}^{} + Z_{B,L}^{\dagger})/2$ to the trivial Potts model $H_{1}$ in practical simulations to explicitly break the $\mathbb{Z}_{3}^{B}$ symmetry; the boundary and bulk magnetizations are measured at sites $1$ and $L/2$, respectively.
} \label{fig:sz}
\end{figure}

\begin{figure}[t]
	\centering
\includegraphics[width=3.5in]{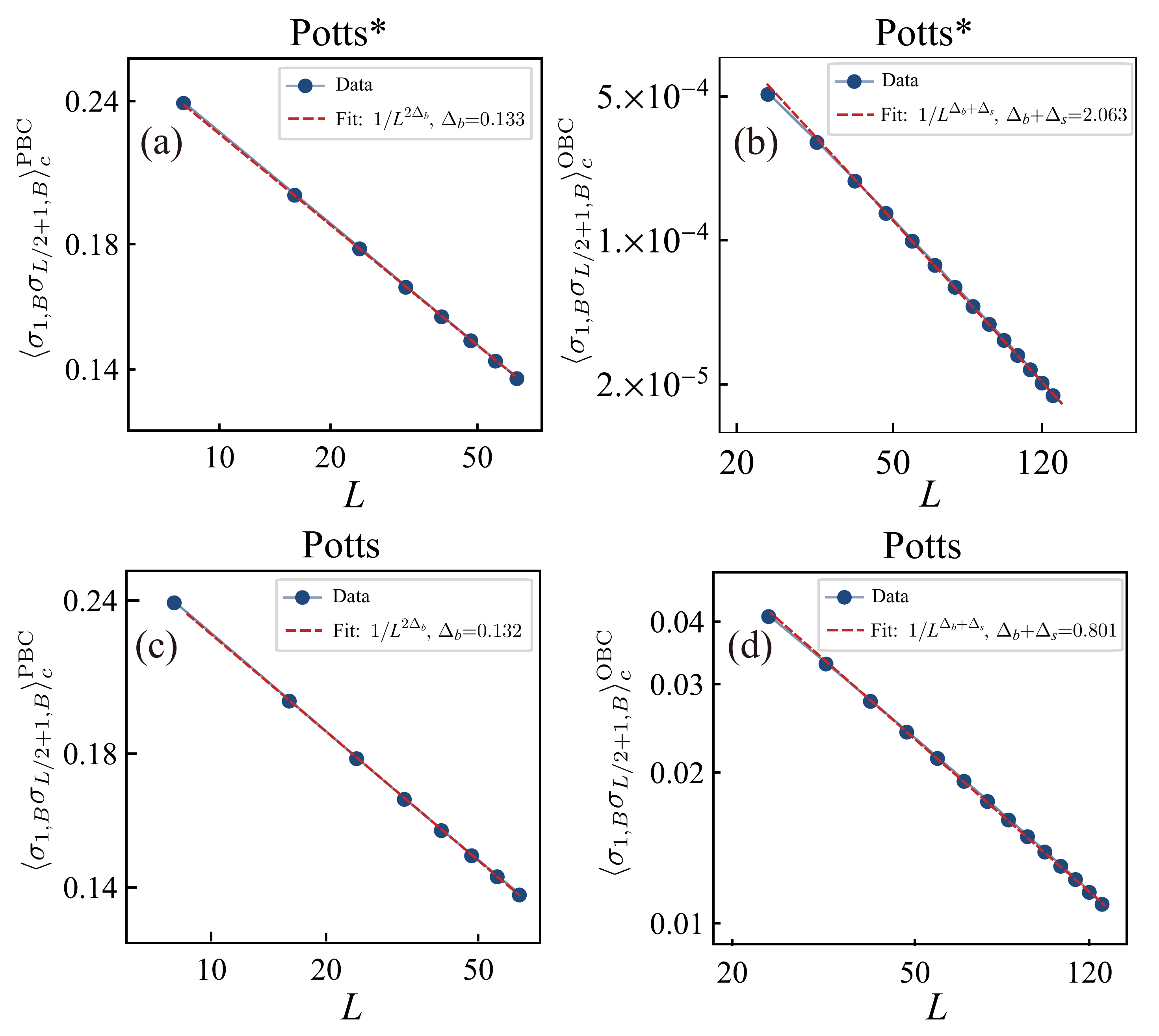}\\
	\caption{Extraction of the bulk and boundary scaling dimensions ($\Delta_{b}$ and $\Delta_{s}$) of the magnetization at topologically distinct Potts QCPs. (a, c) Bulk-bulk connected correlation functions evaluated under periodic boundary conditions for the (a) topologically nontrivial (Potts*) and (c) trivial (Potts) QCPs. Algebraic fits to $1/L^{2\Delta_{b}}$ yield the bulk scaling dimension $\Delta_{b} \approx 0.133$, consistent with the analytic value $2/15$. (b, d) Boundary-bulk connected correlation functions  evaluated under open boundary conditions for the (b) Potts* and (d) Potts QCPs. By fitting the finite-size scaling behaviors, the boundary scaling dimension is extracted as $\Delta_{s} \approx 1.94$ for the nontrivial Potts* QCP, and $\Delta_{s} = 2/3$ for the trivial Potts QCP. Inset of (b) shows the boundary magnetization as a function of system size. Note that $\sigma_{l,B} \equiv (Z_{l,B}^{} + Z_{l,B}^{\dagger}) / 2$.} \label{pott_delta}
\end{figure}

\emph{\textbf{Extracting order-parameter critical exponents in equilibrium.}}---To establish the correct dynamical scaling of the order parameters discussed in the main text, we need to extract the critical exponents \( \beta_{b/s} \) for the bulk and boundary order parameters as a reference. Specifically, we compute the local magnetization as a function of the dimensionless distance to criticality, \( \epsilon \), and employ the equilibrium scaling relation near the critical point, \( m_a \sim \epsilon^{\beta_a} \) where \( a = b,s \) denotes bulk and surface quantities, respectively. As shown in Fig.~\ref{fig:sz}(a)-(d), for a system size \( L = 256 \), both local magnetizations exhibit clear power-law behavior. From this data, we extract \( \beta_b = \frac{1}{8} \) and \( \beta_s = \frac{1}{2} \) for the TFI chain, and \( \beta_b = \frac{1}{8} \) and \( \beta_s = 1 \) for the CI chain. 
In Fig.~\ref{fig:sz}(e)-(h), we further show the local magnetization as a function of the dimensionless distance $\epsilon$ from topologically distinct Potts QCPs. It is noted that 
we have explicitly introduced an additional $B$ site at the left open boundary (which can be labeled as site $0$, with corresponding operators $X_{B,0}^{}$ and $Z_{B,0}^{}$) in our practical calculations for $H_{P2}$ of the main text. This setup not only avoids the spurious degeneracy caused by $[Z_{A,1}^{}, H_{P2}^{}] = 0$, but also renders the entire system spatially more symmetric. Consequently, for both models $H_{P1}^{}$ and $H_{P2}^{}$, the boundary and bulk magnetizations can be formulated in a unified way: $m_{s} \equiv \langle (Z_{B,1}^{} + Z_{B,1}^{\dagger})\rangle / 2$ and $m_{b} \equiv \langle(Z_{B,L/2}^{} + Z_{B,L/2}^{\dagger})\rangle / 2$, respectively. Similar to the CI chain, the boundary magnetization of $H_{P2}$ is extracted from the next-to-boundary $B$ site. The ground states are obtained from density-matrix renormalization group (DMRG) simulations with the bond dimension of the matrix product state (MPS) kept large enough to bound truncation error of the singular value decomposition (SVD) below $10^{-10}$. From the data obtained by DMRG simulations, we extract \( \beta_b =\frac{1}{9} \) and \( \beta_s = \frac{5}{9} \) for the topologically trivial Potts (Potts) QCP, and \( \beta_b = \frac{1}{9} \) and \( \beta_s \approx0.952 \) for the topologically nontrivial Potts (Potts*) QCP. Additionally, as shown in Fig.~\ref{pott_delta}, we extract the bulk and boundary scaling dimensions of the magnetization at topologically distinct Potts QCPs: for the topologically trivial QCP, $\Delta_b = 2/15$ and $\Delta_s = 2/3$; for the topologically nontrivial QCP, $\Delta_b = 2/15$ and $\Delta_s \approx 1.94$\,.

\begin{figure}[tb]
	\centering
\includegraphics[width=3.5in]{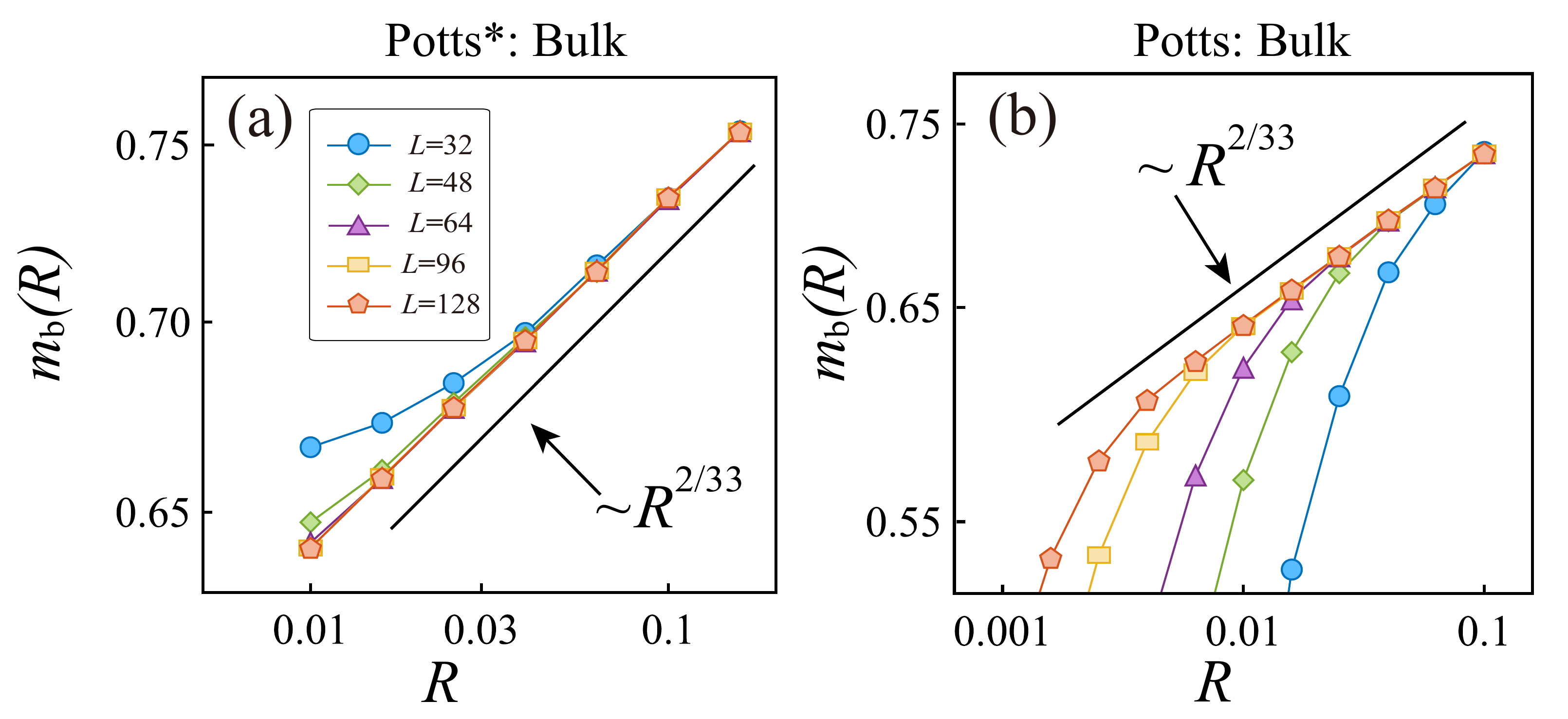}\\
	\caption{The bulk magnetization $m_b(R)$ at the end of the quench  as a function of the quench rate $R$ for (a) the Potts* QCP and (b) the Potts QCP. Different symbols correspond to different system sizes $L$. The dynamical scaling of $m_b(R)$ at topologically distinct QCPs exhibits the same power-law behavior, with an exponent close to $2/33$ (solid lines). }
\label{fig:bulk1}
\end{figure}

\emph{\textbf{Driven critivcal dynamics of  bulk magnetization at topologically distinct Potts QCPs.}}---For completeness, we supplement here the driven critical dynamics of the bulk magnetization at topologically distinct Potts QCPs. As shown in Fig.~\ref{fig:bulk1}, the bulk magnetization exhibits power-law scaling with exponent $2/33$ at both (a) the Potts* QCP and (b) Potts QCP, consistent with the conventional KZ prediction [Eq.~(3) in the main text] using $\beta_b = 1/9$, $z=1$, and $\nu = 5/6$ [see Fig.~\ref{fig:sz}(g) and (h) for the extraction of $\beta_b$] for the (1+1)-dimensional Potts universality class. The dynamics are simulated using the time-dependent variational principle (TDVP) algorithm based on MPS with a discrete time step of $\text{d}t = 0.01$\,. For the topologically trivial model $H_{P1}$, since the $A$ and $B$ degrees of freedom are completely decoupled, we can omit the $A$ sublattice. In this case, we employ the two-site TDVP and strictly keep the SVD truncation error below $10^{-10}$ throughout the evolution. For the topologically nontrivial case, we simulate the full system and, consistent with the equilibrium calculations, include the extra site $0$ for $B$ degree of freedom at the left open boundary. Here, instead of enforcing a stringent SVD truncation error---which would lead to a very large MPS bond dimension during the evolution---we impose a maximum MPS bond dimension $\chi$ and find that the results are well-converged with $\chi = 32$. We note that while the main trend shown in Fig.~3(a) has well converged, the minor fluctuations of $\mathcal{O}(10^{-4})$ superimposed on it cannot be eliminated by increasing the bond dimension $\chi$. These likely originate from the time discretization error and the continuous projection of the real-time dynamics onto a fixed-dimension tangent-space manifold. 

\emph{\textbf{Data collapse of order-parameter dynamics at topologically distinct Ising and Potts QCPs.}}---In Fig.~2 of the main text, we present the scaling behavior of the bulk and boundary local magnetizations as the system is driven from the ferromagnetic phase toward critical points $\lambda_c$ with distinct topological properties. At $\lambda_c$, the magnetization obeys the scaling form
\begin{equation}
\label{eq:sres1}
m_a(R,L)=L^{-\Delta_a} F_a(RL^r), \qquad a=b,s ,
\end{equation}
where $F_a$ is a nonuniversal scaling function.
For the bulk magnetizations of both models and the boundary magnetization of the TFI model, Eq.~\eqref{eq:sres1} reduces to the conventional finite-size scaling form~\cite{Liu2014PRB,Huang2014PRB}: $m_a(R,L)=L^{-\beta_a/\nu} F_a(RL^r), a=b,s$.
For relatively large quench rates $R$, the magnetization exhibits a power-law dependence on $R$ with an exponent close to $\Delta_a/r$, applicable to both bulk and boundary magnetizations in the TFI and CI models. Here $\Delta_a$ is the scaling dimension of the order parameter, and $r=z+1/\nu$ is determined solely by the bulk universality class. Notably, the exponent associated with $R$ is nearly independent of the system size $L$. In contrast, in the slow-quench regime, $m_a(R)$ becomes essentially independent of $R$, and satisfy the finite-size scaling $m_a \propto L^{-\Delta_a}$. 

\begin{figure}[t]
	\centering
\includegraphics[width=7in]{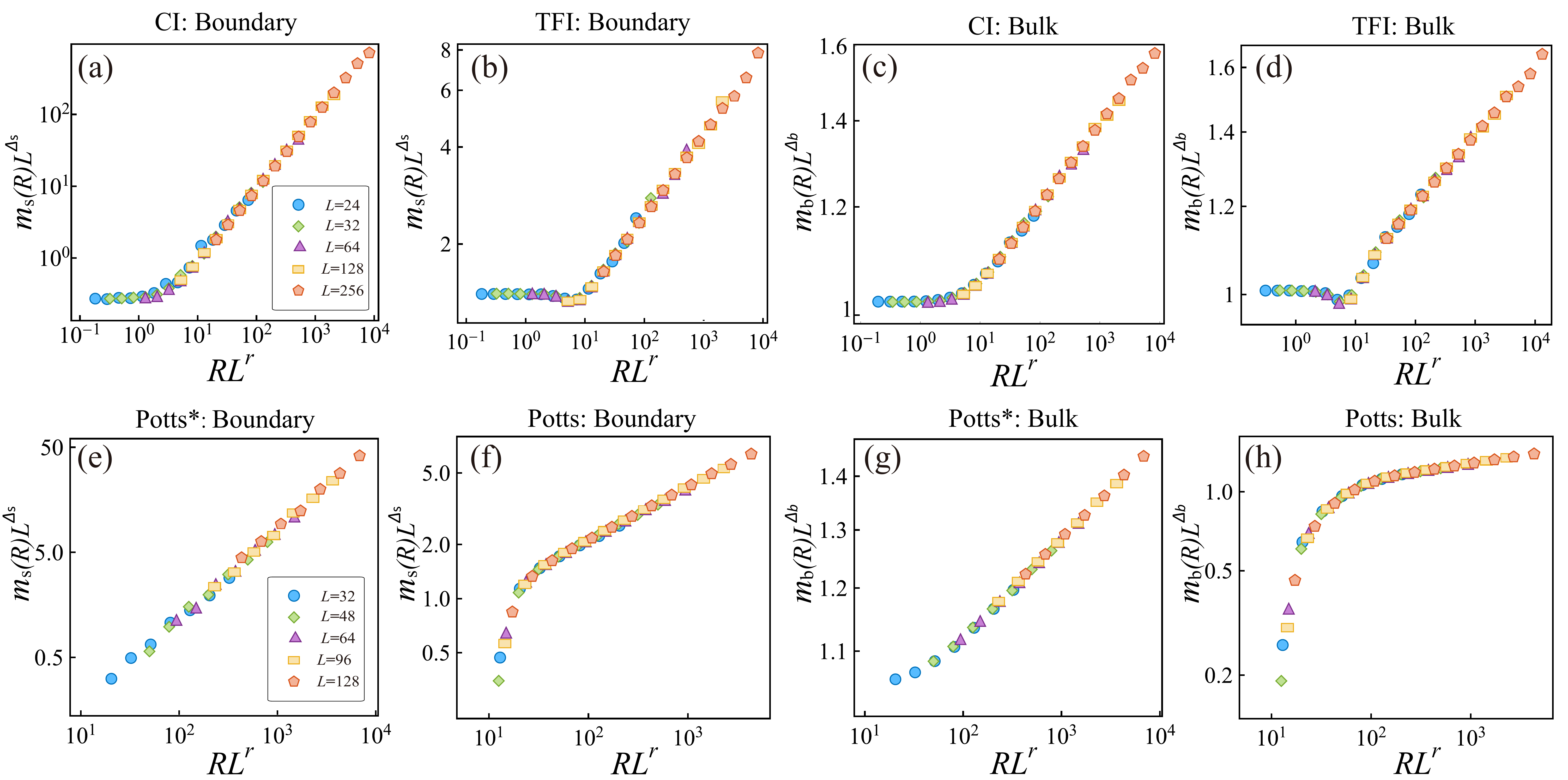}\\
	\caption{The rescaled magnetization $m_a(R)L^{\Delta_a}$ as a function  of $RL^r$ for different $L$. (a)-(d) correspond to panels (a)-(d) in Fig.~2 of the main text. The plots collapse onto a single scaling function, consistent with the dynamical scaling hypothesis [Eq.~\eqref{eq:sres1}] and the following scaling exponents: (a) $ \Delta_s = 2, r = 2 $, (b) $ \Delta_s = 1/2, r = 2 $, (c) $ \Delta_b = 1/8, r = 2 $, and (d) $ \Delta_b = 1/8, r = 2 $. (e)-(h) Same as (a)-(d) but for topologically distinct Potts QCPs.  The plots also collapse onto a single scaling function, consistent with the dynamical scaling hypothesis [Eq.~\eqref{eq:sres1}] and the following scaling exponents: (e) $ \Delta_s = 1.94, r = 11/5 $, (f) $ \Delta_s =2/3 , r = 11/5 $, (g) $ \Delta_b = 2/15, r = 11/5 $, and (h) $ \Delta_b =2/15 , r = 11/5$. All plots are displayed on log-log scales.
} \label{fig:s1}
\end{figure}

Taken together, the dynamical scaling of both order parameters is well captured by Eq.~\eqref{eq:sres1}. As shown in Fig.~\ref{fig:s1}, all data collapse onto a single universal curve when $m_a(R)$ and $R$ are rescaled as $m_a(R)L^{\Delta_a}$ and $RL^r$, respectively, using the corresponding scaling exponents, demonstrating universal dynamical scaling of both bulk and boundary order parameters at topologically distinct Ising QCPs. To further verify the generality of Eq.~\eqref{eq:sres1} beyond exactly solvable models, we consider genuinely interacting quantum Potts chains that cannot be reduced to a free-fermion description~\cite{Alica2016ARCMP}. For topologically distinct Potts  QCPs, the numerical results are shown in Fig.~\ref{fig:s1}(e)-(h). It can be seen that when using the corresponding scaling exponents to rescale $m_a(R)$ and $R$ as $m_a(R)L^{\Delta_a}$ and $RL^r$, respectively, all data points collapse onto a single universal curve.

\emph{\textbf{Brief introduction to the $\alpha$-chain and its gSPT physics.}}
To illustrate the topological physics more intuitively, we reformulate the fermionic models in terms of Majorana operators. Specifically, we consider a noninteracting $\alpha$-neighbor Hamiltonian on a one-dimensional Majorana chain~\cite{Verresen2018PRL,verresen2020topologyedgestatessurvive}:
\begin{equation}
H_{\alpha} = i \sum_{n} \gamma_{2n} \gamma_{2(n+\alpha)-1},
\end{equation}
where $\gamma_{2n-1}$ and $\gamma_{2n}$ represent two Majorana species within each site. For $\alpha = 1$, $H_{\alpha}$ reduces to the Kitaev chain, which hosts a single Majorana zero mode at each edge. In general, $H_{\alpha}$ supports $|\alpha|$ Majorana zero modes per edge and can be viewed as a stack of $|\alpha|$ Kitaev chains. Consequently, its ground state corresponds to a gapped topological phase with a $2{|\alpha|}$-fold degenerate edge mode.

Topological critical Hamiltonians can be constructed as linear combinations of different $\alpha$-chains, leading to continuous phase transitions between gapped phases with distinct topological invariants. Importantly, whenever such a transition involves a change in a nonzero topological index, the resulting QCP hosts exponentially localized Majorana edge modes~\cite{Verresen2018PRL,YU20261}. This provides a general guiding principle for engineering gSPT physics in free-fermion tight-binding models.

To be concrete, we consider the following one-dimensional free fermion Hamiltonian that exhibits topologically distinct QCPs:
\begin{equation}
\label{eq:s21}
H_{\text{F}} = g_0 H_0 - g_1 H_1 + g_2 H_2,
\end{equation}
where $g_0$, $g_1$, and $g_2$ denote onsite, nearest-neighbor, and next-nearest-neighbor couplings, respectively. Without loss of generality, we fix $g_0 + g_1 + g_2 = 4$. For $g_2 = 0$, the model undergoes a continuous phase transition between a topological superconducting phase with winding number $\omega = 1$ and a trivial phase at $g_0 = g_1$. This case corresponds to the one-dimensional Kitaev model, whose Jordan-Wigner dual is the transverse-field Ising chain.

In contrast, for $g_0 = 0$, the model exhibits a continuous transition between topological superconducting phases with $\omega = 1$ and $\omega = 2$. This QCP host exponentially localized and two-fold degenerate Majorana edge modes~\cite{Verresen2018PRL,verresen2020topologyedgestatessurvive}, thus constituting topologically nontrivial critical points. Their Jordan-Wigner dual corresponds to the cluster-Ising model, where the critical points map to symmetry-enriched Ising criticality~\cite{Verresen2021PRX,Yu2022PRL}.

\begin{figure}[b]
	\centering
\includegraphics[width=7in]{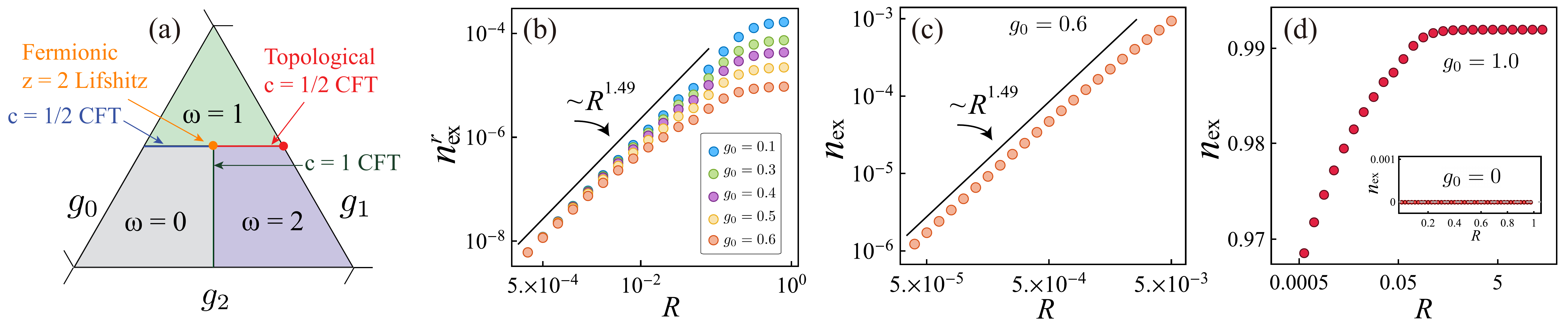}\\
	\caption{ (a) Schematic phase diagram of $ H_{\text{F}} $. The phase boundaries (solid colored lines) denote quantum critical lines described by conformal field theories with central charge $c$. The multicritical point (orange dot) serves as the quantum critical point of a continuous transition between two topologically distinct $c=1/2$ CFTs, represented by the red and blue lines, respectively. 
(b) Rescaled edge-excitation  as a function of the quench rate $R$ for different values of $g_0$. In the slow-quench regime, the edge excitations exhibit a power-law scaling with an exponent close to $1.5$. As $g_0$ increases, this power-law behavior progressively breaks down at larger quench rates. (c) For $g_0 = 0.6$, the power-law scaling of the edge excitations is recovered within a narrower range of quench rates.
(d) In the limiting case $g_0 = 0$, the edge states are completely decoupled from the bulk, and no edge excitations are generated (see inset). By contrast, for $g_0 = 1$, the number of excitations  approaches unity, reflecting the delocalization of the topological edge modes into the gapless bulk.
 } \label{fig:s2}
\end{figure}

For generic values of $g_0$, $g_1$, and $g_2$, the resulting global phase diagram (Fig.~\ref{fig:s2}(a)) contains both topologically trivial (blue) and nontrivial (red) quantum critical lines, which intersect at a multicritical point (orange dot at $g_0 = 1$), representing a transition between topologically distinct classes of quantum criticality.

\emph{\textbf{Effect of bulk coupling on dynamical scaling at a topologically nontrivial QCP. }}
In this subsection, we examine the effect of bulk–edge coupling on dynamical scaling at a topologically nontrivial QCP. Without loss of generality, we consider the Hamiltonian introduced in Eq.~(\ref{eq:s21}). The corresponding phase diagram, shown in Fig.~\ref{fig:s2} (a), contains three distinct topological phases characterized by the invariants $\omega = 0, 1, 2$. The phase boundaries denote quantum critical lines described by conformal field theories with central charge $c$. The multicritical point (orange dot) is the only nonconformal critical point and is described by a fermionic Lifshitz theory with dynamical critical exponent $z = 2$. It serves as the QCP of a continuous phase transition between two topologically distinct $c = 1/2$ CFTs, represented by the red and blue lines.


 We linearly vary $g_1$ and $g_2$ according to:
\begin{equation}
\begin{cases}
g_1(t) = g_{1,i} + (g_{1,c} - g_{1,i})Rt, \\
g_2(t) = 4 - g_0 - g_1(t),\quad t\in[0,1/R],
\end{cases}
\end{equation}
such that the constraint $ g_0 + g_1(t) + g_2(t) = 4$ is maintained throughout the evolution. The ramp starts from the ground state in the topological superconducting phase with $\omega = 1$ and terminates on the topologically nontrivial quantum critical line (red line). The parameter $g_0$ is set to various values, characterizing the strength of the bulk-edge coupling.

In Fig.~\ref{fig:s2} (b), we show the edge-excitation  as a function of the quench rate $R$ for different values of $g_0$. To facilitate a direct comparison of the power-law behavior across different $g_0$, we rescale the number of excitations  for each $g_0$ by a constant normalization factor: $n_{ex}^r(g_0) = \frac{n_{ex}(g_0) \cdot n_{ex}(g_0 = 0.1, R = 10^{-3.5})}{n_{ex}(g_0, R = 10^{-3.5})}.$ For small values of $g_0$, the number of edge-excitations  exhibits a clear power-law scaling $ n_{ex} \propto R^{1.49}$. As $g_0$ increases, this power-law behavior gradually breaks down at relatively large quench rates and survives only in the slow-quench regime. 

To assess whether the power-law scaling of excitations persists for larger values of $g_0$ in the adiabatic limit of small quench rates $R$, we perform simulations with even slower ramps and evaluate the resulting excitation production for $g_0 = 0.6$, as shown in Fig.~\ref{fig:s2} (c). The results confirm that the power-law behavior is recovered in the slow-quench regime.

We further examine two limiting cases, $g_0 = 0$ and $g_0 = 1$. For $g_0 = 0$, even at the topologically nontrivial QCP, the topological edge modes remain completely decoupled from the bulk degrees of freedom . As a result, no edge excitations are generated throughout the quench protocol, as shown in the inset of Fig.~\ref{fig:s2} (d). By contrast, for $g_0 = 1$, where the quench terminates at the multicritical point, the topological edge modes delocalize into the gapless bulk. Consequently, the excitation production no longer exhibits power-law scaling, as illustrated in Fig.~\ref{fig:s2} (d).

In summary, our results unambiguously demonstrate that when $g_0$ is smaller than a threshold value $1.0$, the topological edge modes persist and coexist with the gapless bulk, leading to \emph{universal} anomalous power-law scaling of $n_{\text{ex}}$ with $R$, characterized by the same exponent $\sim 1.5$. In contrast, when $g_0$ exceeds $1.0$, the edge modes  delocalize and merge into the gapless bulk, and the dynamical scaling reverts to the usual behavior expected at trivial QCPs, $n_{\text{ex}} \sim 1.0$. 

\begin{figure}[b]
	\centering
\includegraphics[width=1.75in]{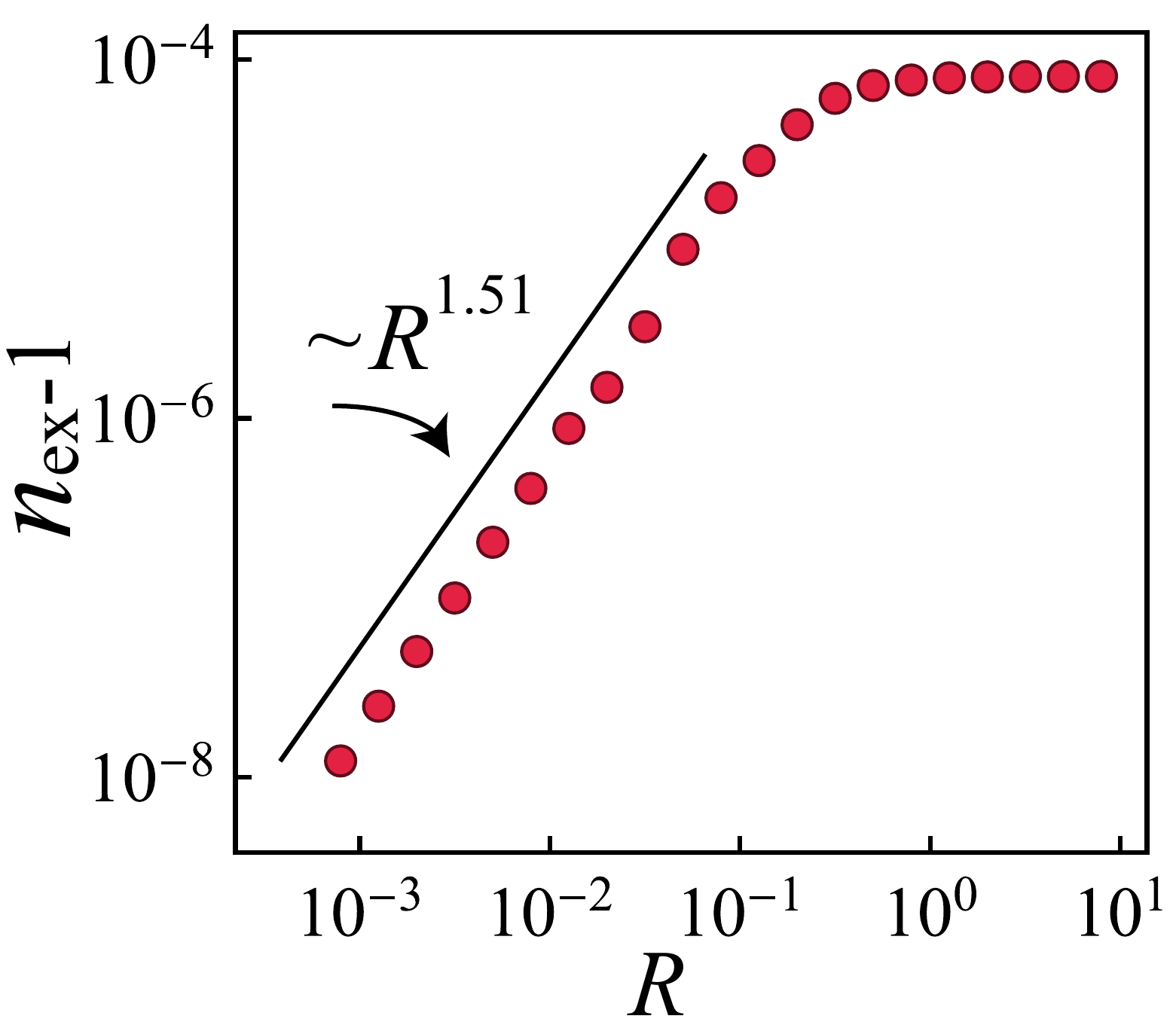}\\
	\caption{When the system is driven from one gapped topological phase across a topologically nontrivial QCP into another gapped topological phase, the number of edge-excitations exhibits the same anomalous dynamical scaling behavior as that reported in the main text.
 } \label{fig:s3}
\end{figure}

\emph{\textbf{ Quench across a topological nontrivial critical point. }}---In this subsection, we investigate the dynamical scaling of excitations when the system is driven from one gapped topological phase, \emph{across} a topologically nontrivial QCP, into another gapped topological phase. We employ a quench protocol consistent with that used in the main text, $g_2(t)=g_{2,i}+(g_{2,f}-g_{2,i})Rt, t\in[0,1/R]$, where $g_{2,i}$ and $g_{2,f}$ correspond to two distinct gapped topological phases. As the quench drives the system from a phase with winding number $\omega=1$ to one with $\omega=2$, a single state from the valence band necessarily merges into the bulk continuum, contributing an additional unit of excitation. After subtracting this trivial contribution, the resulting excitation number is shown in Fig.~\ref{fig:s3}. In the slow-quench regime, the excitations display a robust power-law dependence on the quench rate $R$, with an exponent close to $1.5$, fully consistent with the anomalous scaling obtained for quenches terminating at a topologically nontrivial QCP.

\begin{figure}[t]
	\centering
\includegraphics[width=5.6in]{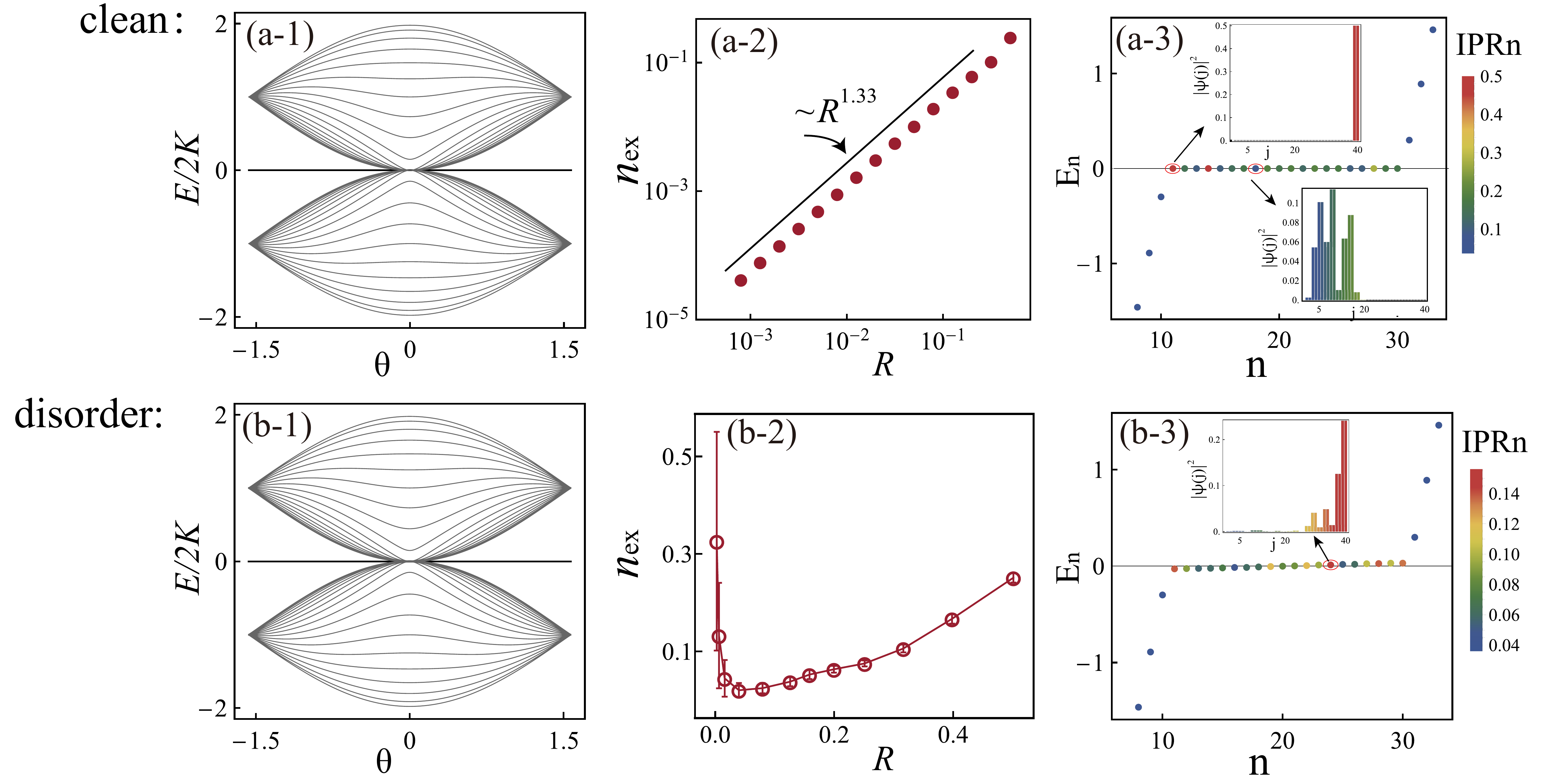}\\
	\caption{(a) Clean system. (a-1) Energy spectrum of the Creutz ladder with open boundary conditions as a function of the magnetic flux $\theta$. Two in-gap zero-energy modes are present, corresponding to localized topological edge states. (a-2) Excitation  generated during a magnetic-flux quench $\theta(t)$, plotted as a function of the quench rate $R$ on a logarithmic scale, for an initial edge state. (a-3) The energy spectrum as a function of eigenvalue index around zero energy at the critical point \( \theta_c \) exhibits multiple degenerate zero-energy levels. The color scale indicates the inverse participation ratio (IPR) of the corresponding eigenstates. As shown in the inset, these zero-energy modes contain both localized edge states (dark red) and extended bulk states (dark blue).
(b) Disordered system. Same as (a), but in the presence of weak disorder. Although the single-particle spectrum remains largely unchanged [see (b-1)], the power-law scaling of the number of excitations is completely suppressed [see (b-2)]. The average is performed over $600$ independent disorder realizations. (b-3) Hybridization between edge states and zero-energy extended bulk states leads to the delocalization of the edge modes into the bulk. Panel (2) was obtained for a system size of $L=400$, whereas for clarity, we set $L=20$ for panels (1) and (3).}
 \label{fig:s5}
\end{figure}

\emph{\textbf{Brief review of the Creutz ladder model.}}---The Creutz ladder model, originally introduced by Creutz~\cite{Creutz1999PRL}, describes spinless fermions moving on a two-leg ladder and is governed by the Hamiltonian
\begin{equation}
\label{SE1}
H_{\text{Creutz}} = -\sum_{l=1}^{L} \Big[ K_1\big(e^{-i\theta} a^{\dagger}_{l+1} a_l + e^{i\theta} b^{\dagger}_{l+1} b_l\big) + K_2\big(b^{\dagger}_{l+1} a_l + a^{\dagger}_{l+1} b_l\big) + Ma^{\dagger}_{l} a_l + \text{H.c.} \Big],
\end{equation}
where $a_l$ ($b_l$) annihilates a fermion on the upper (lower) leg at rung $l$~\cite{Creutz1999PRL}. The horizontal hopping amplitude is $K_1$, the diagonal hopping is $K_2$, and $M$ denotes the vertical coupling; all parameters are taken to be positive. Following Ref.~\cite{Bermudez2009PRL}, we set $K_1 = K_2 \equiv K $ and $M = 0 $ in the subsequent analysis.

An external perpendicular magnetic field introduces a flux $\theta\in[-\pi/2,\pi/2]$ per plaquette, giving rise to quantum interference that confines fermions to the ladder edges and produces topological edge states. For weak vertical coupling, $M<K$, the system undergoes a continuous quantum phase transition at $\theta_c=0$, characterized by the equilibrium critical exponents $\nu=z=1$. In the following, we investigate the quench dynamics of the edge states in both clean and disordered settings.

In the clean system, we prepare an initial zero-energy state at $\theta=-\pi/2$, which is exponentially localized at the edge. We then linearly ramp the magnetic flux according to $\theta(t)=Rt-\pi/2$, with $\theta\in[-\pi/2,\pi/2]$, and compute the total excitation number at the end of the quench,$n_{\mathrm{ex}}=\sum_{E_n>0}\big|\langle E_n(t_f)|\Psi(t_f)\rangle\big|^2$, where $|E_n(t_f)\rangle$ denotes the eigenstates of the final Hamiltonian and $|\Psi(t_f)\rangle$ is the time-evolved state obtained by time-evolving an edge state of the initial Hamiltonian. As shown in Fig.~\ref{fig:s5} (a–2), the numerical results reveal a pronounced deviation from standard Kibble–Zurek scaling, with the number of excitations following an anomalous power law $n_{\mathrm{ex}}\propto R^{1.33}$. This behavior is consistent with topology-induced anomalous defect production reported in the literature~\cite{Bermudez2009PRL}.

To examine the robustness of the anomalous dynamics of defect production in the Creutz ladder model against disorder, we introduce randomness in the hopping amplitudes $K_{1,2}^{\mathrm{dis}}$, which are independently drawn at each site from a uniform distribution $K_{1,2}^{\mathrm{dis}}\in K_{1,2}[1-\delta,,1+\delta]$. As shown in Fig.~3 (b–3) of the main text, for a moderate disorder strength $\delta=0.1$, the power-law scaling of the number of excitations  is completely suppressed under the  quench protocol, $\theta(t)=Rt-\pi/2$ with $\theta\in[-\pi/2,\pi/2]$. We further consider a weaker disorder, $\delta=0.01$. Although the single-particle spectra of the clean and disordered systems remain nearly identical [see Figs.~\ref{fig:s5} (a–1) and \ref{fig:s5} (b–1)], the power-law scaling of excitations is nevertheless fully destroyed in the presence of disorder [cf. Figs.~\ref{fig:s5} (a–2) and \ref{fig:s5} (b–2)].

To elucidate the origin of this breakdown, we analyze the energy spectrum at the critical point $\theta_c$ for both the clean and disordered cases, as shown in Figs.~\ref{fig:s5} (a–3) and \ref{fig:s5} (b–3). In the clean system, multiple states are degenerate at zero energy. The color coding indicates the inverse participation ratio, ${\rm IPR}_n=\sum_{j=1}^{L}|\psi_n(j)|^4$, of the corresponding eigenstate $|\psi_n\rangle$. As illustrated in the inset of Fig.~\ref{fig:s5} (a–3), these degenerate zero energy modes contains both exponentially localized edge states (corresponding to the dark red points) and extended bulk states (corresponding to dark blue points). Thus, even weak disorder induces hybridization between the edge states and the zero-energy bulk states, causing the edge states to delocalize into the bulk, as shown in the inset of Fig.~\ref{fig:s5} (b–3).  This ultimately leads to the disappearance of the anomalous power-law scaling of the number of excitations  in the disordered Creutz ladder.



\begin{figure}[t]
	\centering
\includegraphics[width=3.5in]{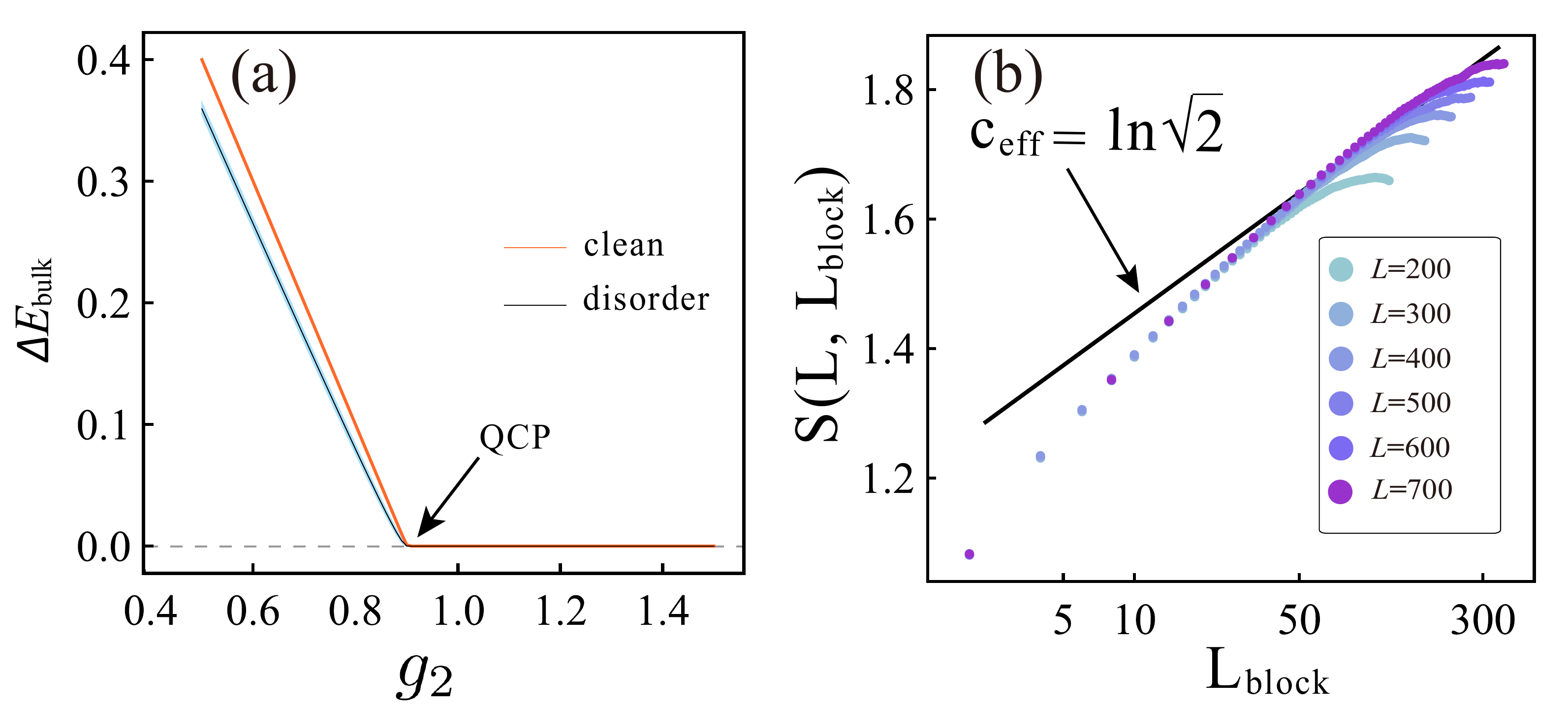}\\
	\caption{(a) Bulk energy gap as a function of $ g_2 $ for the clean (orange solid line) and disordered (black solid line) Hamiltonian. In both cases, the gap closes at the same critical value $ g_{2,c}$. The bule shaded region denotes one standard deviation above and below the disorder-averaged gap. (b) Entanglement-entropy scaling consistent with an infinite-randomness fixed point characterized by $ c_{\mathrm{eff}} = \ln \sqrt{2} $ (black lines are guides to the eye). All data are averaged over $600$ independent disorder realizations.}
\label{fig:s6}
\end{figure}

\emph{\textbf{Entanglement scaling at disordered topologically nontrivial QCPs: }} We first consider the clean limit of the Hamiltonian $H = \sum_{\alpha=0}^{2} g_{\alpha} H_{\alpha}$. To locate the critical point, we fix $g_0 = 0.1$ and $g_1 = 1$ and plot the bulk energy gap as a function of $ g_2$, as shown in Fig.~\ref{fig:s6} (a). The results show that the bulk gap closes at $ g_{2,c} = 0.9$, indicated by the orange solid line. We further introduce symmetry-preserving disorder through $g_{0/1}^{\text{dis}}$, which is drawn independently at each site from a uniform distribution given by $g_{0/1}^{\text{dis}} \in g_{0/1}[1-\delta, 1+\delta]$, where the disorder strength is $\delta = 0.1$. As depicted in Fig.~\ref{fig:s6} (a), the bulk energy gap $\Delta E$ of the disordered system also closes at $g_{2,c} = 0.9$ (black solid line).

We next examine whether the system flows to an infinite-randomness fixed point, characterized by an effective central charge $c_{\mathrm{eff}} = \ln \sqrt{2}$. To this end, we diagonalize periodic systems of size $L$ and compute the entanglement entropy $S(L, L_{\mathrm{block}})$ for a subsystem of length $L_{\mathrm{block}}$. The method for calculating entanglement entropy is presented in Sec.~\ref{sec:1}. After disorder averaging, the entanglement entropy is expected to obey the asymptotic logarithmic scaling form $S \sim (c_{\mathrm{eff}}/3) \ln L_{\mathrm{block}}$ for $1 \ll L_{\mathrm{block}} \ll L/2$, as illustrated in Fig.~\ref{fig:s6} (b).

\section{Detailed explanation of the two definitions of the number of excitations}
\label{sec:2.5}

\emph{\textbf{Definition of the excitation number in the main text: }}The quantity $n_{\text{ex}}$ introduced in the main text represents the number of excitations and is a commonly used observable in studies of driven dynamics in free-fermion systems~\cite{Ulifmmode2019PRB,Liou2018PRB,Deng2025PRL}. To avoid confusion, we refer to this dimensionless quantity as the “number of excitations”. Next, we provide a detailed explanation of the quench protocol and the definition of the excitation number.

\begin{figure}[t]\label{fig:eigsta1}
	\centering
\includegraphics[width=3.25in]{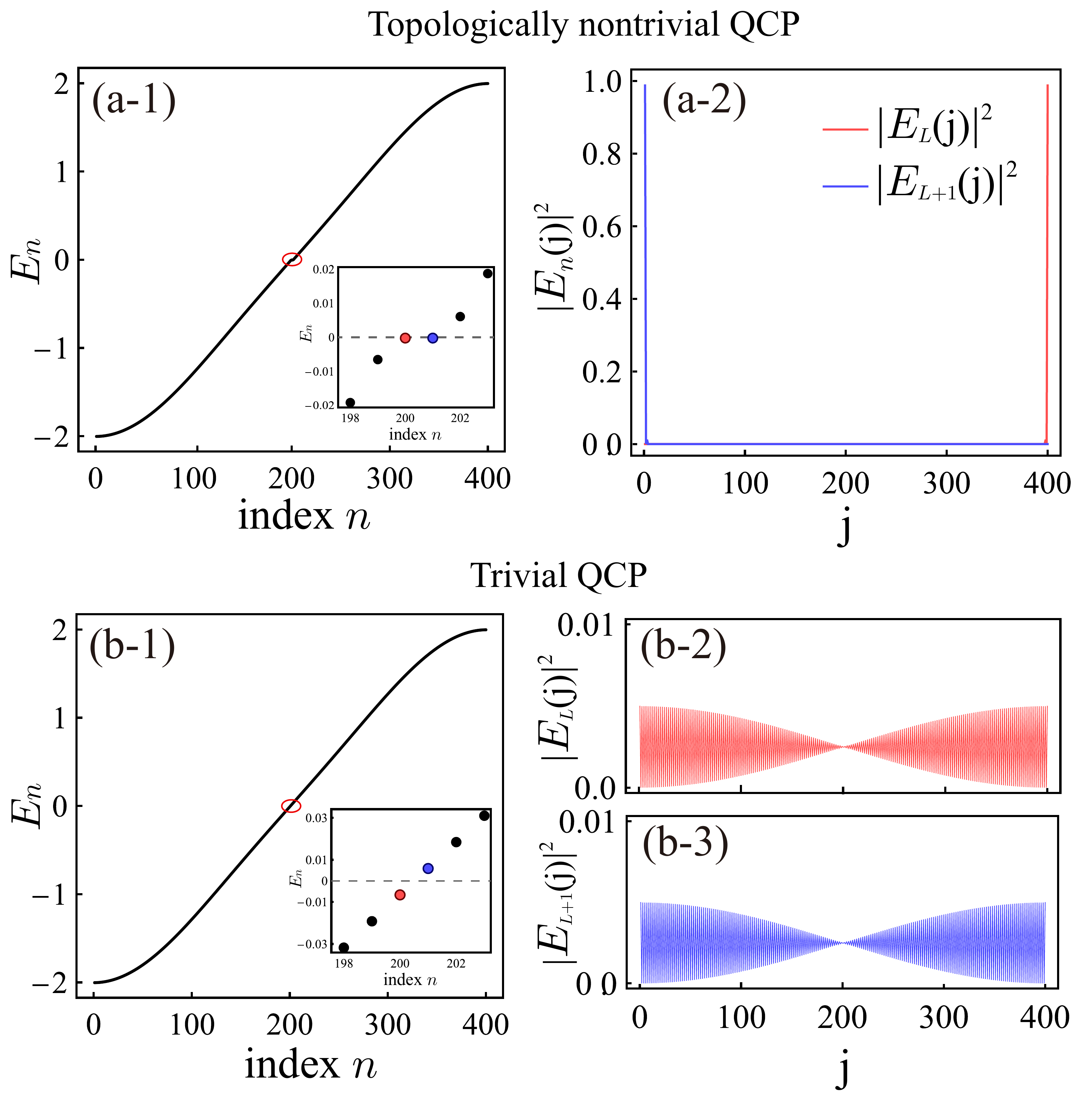}\\
	\caption{ 
    Energy spectrum of the free-fermion Hamiltonian  as a function of the state index at (a-1) a topologically nontrivial QCP and (b-1) a trivial QCP. The insets show a magnified view of the intermediate energy levels. At the topologically nontrivial critical point, a pair of degenerate zero-energy states $|E_L\rangle$ and $|E_{L+1}\rangle$ appears, whose probability distributions are shown in (a-2), exhibiting localization at the boundaries. In contrast, at the trivial critical point, the states $|E_L\rangle$ and $|E_{L+1}\rangle$ are bulk states and do not exhibit edge localization, as shown in (b-2) and (b-3). All calculations are performed under open boundary conditions, with $L$ denoting the number of unit cells and a total system size of $2L=400$.
} \label{energyspec}
\end{figure}

\emph{\textbf{Specific Example: }}Specifically, we consider the following one-dimensional free-fermion Hamiltonian exhibiting topologically distinct QCPs:
\begin{equation}
\label{eq:s21}
H = g_0 H_0 - g_1 H_1 + g_2 H_2,
\end{equation}
where $H_{\alpha} = i \sum_{n} \gamma_{2n} \gamma_{2(n+\alpha)-1}) ((\alpha = 0,1,2)$ describes different noninteracting $\alpha$ chains~\cite{Verresen2017PRB,Verresen2018PRL}. Here, $\gamma_{2n-1}$ and $\gamma_{2n}$ denote two species Majorana modes within each unit cell, and $g_0$, $g_1$, and $g_2$ correspond to onsite, nearest-neighbor, and next-nearest-neighbor couplings, respectively.

We first describe the quench protocol that drives the system to a topologically nontrivial QCP. We fix $g_0 = 0.1$ and $g_1 = 1$, and ramp $g_2$ linearly as $g_2(t) = g_{2,i} + (g_{2,c} - g_{2,i}) R t$, with $t \in [0, 1/R]$. This protocol realizes a quantum phase transition (QPT) between gapped phases with topological invariants $\omega = 1$ and $\omega = 2$. The evolution starts from the gapped phase with $\omega = 1$ at $t_i = 0$ and ends at the topologically nontrivial QCP at $t_f = 1/R$.

\begin{figure}[t]
	\centering
\includegraphics[width=7.5in]{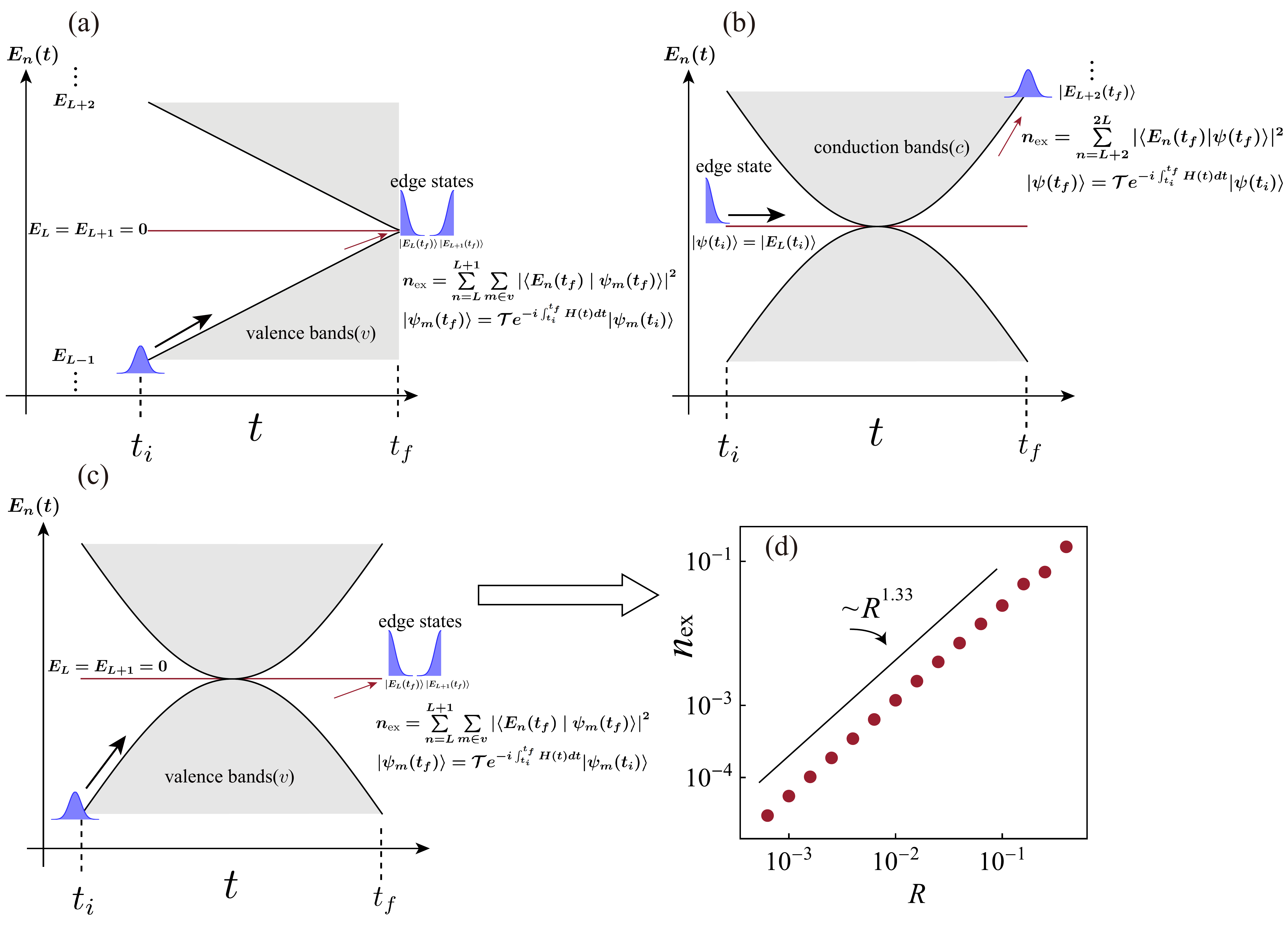}\\
	\caption{
    Schematic illustration of the definition of the number of excitations. (a) Definition used in the main text [Eq.~(\ref{eq:rpex})]. The initial valence-band eigenstates $\{|E_{m}(t_i)\rangle, m\in v=(1,...,L-1)\}$ are fully occupied, i.e., $|\psi_m(t_i)\rangle = |E_m(t_i)\rangle$ , where $v$ denote the valence bands. The system is then driven to a topologically nontrivial quantum critical point at $t_f$, and the time-evolved states $\{|\psi_m(t_f)\rangle, m\in v\}$ are projected onto the eigenstates $|E_L(t_f)\rangle$ and $|E_{L+1}(t_f)\rangle$ of the final Hamiltonian $H(t_f)$, with the contributions summed.  At the topologically nontrivial QCP, $|E_L(t_f)\rangle$ and $|E_{L+1}(t_f)\rangle$ correspond to edge states [see Fig.~3\ref{fig:eigsta1}(a-2)]. (b) Definition used in the Sec.~II [Eq.~(\ref{eq:rpex2})] was proposed in Ref.~\cite{Bermudez_2010} to study the generation of excitations  in the Creutz ladder model. Initially, the zero-energy edge state $|E_L(t_i)\rangle$ is occupied, i.e., $|\psi(t_i)\rangle = |E_L(t_i)\rangle$. The system is then driven through the critical point. At the final time $t_f$, the time-evolved state $|\psi(t_f)\rangle$ is projected onto excited state $|E_n(t_f)\rangle$ with $E_n(t_f)>0$, and the contributions are summed.  (c) Number of excitations $n_{\rm ex}$ defined by Eq.~(\ref{eq:rpex}) in the Creutz ladder model, and (d) $n_{\rm ex}$ as a function of the quench rate $R$, exhibit the same scaling behavior as that defined by Eq.~(\ref{eq:rpex2}) (see Ref.~\cite{Bermudez_2010}). However, the scaling behavior obtained from both definitions is unstable in the  Creutz ladder model with symmetry-preserving disorder. For clarity, only the relevant transitions are indicated by red arrows; other transitions are omitted.} \label{nex}
\end{figure}

To characterize the dynamical scaling behavior associated with topological edge states at criticality, we define the number of excitations as
\begin{equation}
\label{eq:rpex}
n_{\text{ex}} = \sum_{n=L}^{L+1} \sum_{m\in v} \left| \langle E_n(t_f) \mid \psi_m(t_f) \rangle \right|^2,
\end{equation}
where $L$ is the system size.  Here, $v$ denotes the valence bands and $\{ |\psi_m(t_f)\rangle, m \in v={(1,...,L-1)} \}$ is the set of states obtained by time-evolving the initially occupied valence-band states of the initial Hamiltonian $H(t_i)$, and $|E_n(t_f)\rangle$ ($n = L, L+1$) denote the edge-localized eigenstates of the final Hamiltonian $H(t_f)$. A schematic illustration of the definition of the number of excitations according to Eq.~(\ref{eq:rpex}) is shown in Fig.~\ref{nex} (a). This choice of $n = L, L+1$ corresponds to the edge-localized eigenstates because, due to chiral symmetry, the topological zero modes are located near the center of the $2L$-level energy spectrum (see Fig.~\ref{energyspec}(a) for illustration). It is worth emphasizing that similar definitions of excitations—often referred to as edge or in-gap excitations—have been studied in systems such as two-dimensional Chern insulators and higher-order topological insulators~\cite{Ulifmmode2019PRB,Liou2018PRB,Deng2025PRL}.

The states $|E_n(t_f)\rangle$ with $n = L, L+1$ denote the instantaneous eigenstates located near the center of the $2L$-level energy spectrum of the Hamiltonian at the final time $t_f$. The states $\{|\psi_{m}(t_f)\rangle,m\in v\}$ are obtained by time-evolving the initial valence-band eigenstates of the Hamiltonian at $t_i$ under the time-dependent Schrödinger equation. Therefore, $|E_n(t_f)\rangle$ and $|\psi_v(t_f)\rangle$ should reside in the same Hilbert space.

In one-dimensional free-fermion critical systems, for topologically nontrivial QCPs that host robust edge states even when the bulk is gapless, the states $|E_n(t_f)\rangle$ with $n = L, L+1$ coincide with the topological zero-energy edge modes due to chiral symmetry, as shown in Fig.~\ref{nex}(a). In contrast, for topologically trivial QCPs, the $|E_n(t_f)\rangle$ with $n = L, L+1$ correspond to bulk states, since the edge modes delocalize and merge into the gapless bulk in this case. 
Consequently, only in the trivial case do both $|E_n(t_f)\rangle$ and $|\psi_v(t_f)\rangle$ belong to bulk states.

For comparison, we also consider a quench protocol that drives the system to a topologically trivial QCP. In this case, we fix $g_1 = 1$ and $g_2 = 0.1$, and ramp $g_0$ linearly according to $g_0(t) = g_{0,i} + (g_{0,c} - g_{0,i}) R t$, with $t \in [0, 1/R]$. This protocol realizes a QPT from a gapped phase with topological invariant $\omega = 1$ to a trivial phase with $\omega = 0$. The evolution starts from the gapped topological phase and ends at the trivial QCP. As the system approaches the trivial critical point, the edge modes delocalize and merge into the gapless bulk. For consistency, we still select the states at the center of the energy spectrum ($n = L, L+1$) in the definition of $n_{\text{ex}}$, although these states are no longer topological edge modes at criticality (see Fig.~\ref{energyspec}(b)) and instead correspond to bulk states. Consequently, the excitation number defined in Eq.~(\ref{eq:rpex}) approaches $n_{\text{ex}} \to 1$. We include it here to further demonstrate that the anomalous dynamical scaling behavior originates from the presence of robust topological edge states at the QCP.

\emph{\textbf{Definition of the excitation number in the Supplemental Material Sec II: }}  The quantity $\sum |\langle E | \psi(t) \rangle|^2$ introduced in the Supplemental Material Sec II follows the definition originally proposed by A. Bermudez et al~\cite{Bermudez_2010}. For clarity, we rewrite it as
\begin{equation}\label{eq:rpex2}
n_{\text{ex}} = \sum_{E_n > 0} |\langle E_n(t_f) \mid \psi(t_f) \rangle|^2,
\end{equation}
where ${|E_n(t_f)\rangle, n \in ({L+2, \dots, 2L}})$ denotes the set of excited eigenstates of the final Hamiltonian, and $|\psi(t_f)\rangle$ is the state obtained by time-evolving an eigenstate of the initial Hamiltonian $H(t_i)$. In their work, the initial state is chosen as the topological edge eigenstate of the initial Hamiltonian $H_{\text{Creutz}}(t_i)$. Accordingly, the excitation number is defined as the projection of this zero-energy edge state onto the excited (conduction-band) states of the final Hamiltonian, as illustrated in Fig.~\ref{nex}(b).

In contrast, in the definition of $n_{\text{ex}}$ used in the main text , the excitation number is obtained by projecting a set of initially occupied valence-band states onto the zero-energy topological edge modes of the final Hamiltonian, as shown in Fig.~\ref{nex}(a). In fact, these two definitions of $n_{\text{ex}}$ yield identical scaling behavior, as demonstrated in Fig.~\ref{nex} (c) and (d), where the definition from the main text  is applied to the Creutz model. However, unlike in our case, the dynamical scaling in the Creutz model~\cite{Creutz1999PRL} is not robust against symmetry-preserving disorder, as explicitly demonstrated in the Supplemental Material Sec II.

\section{Universal dynamical scaling at the critical points of general $\alpha$ chains}
\label{sec:3}
\begin{figure}[b]
	\centering
\includegraphics[width=5.35in]{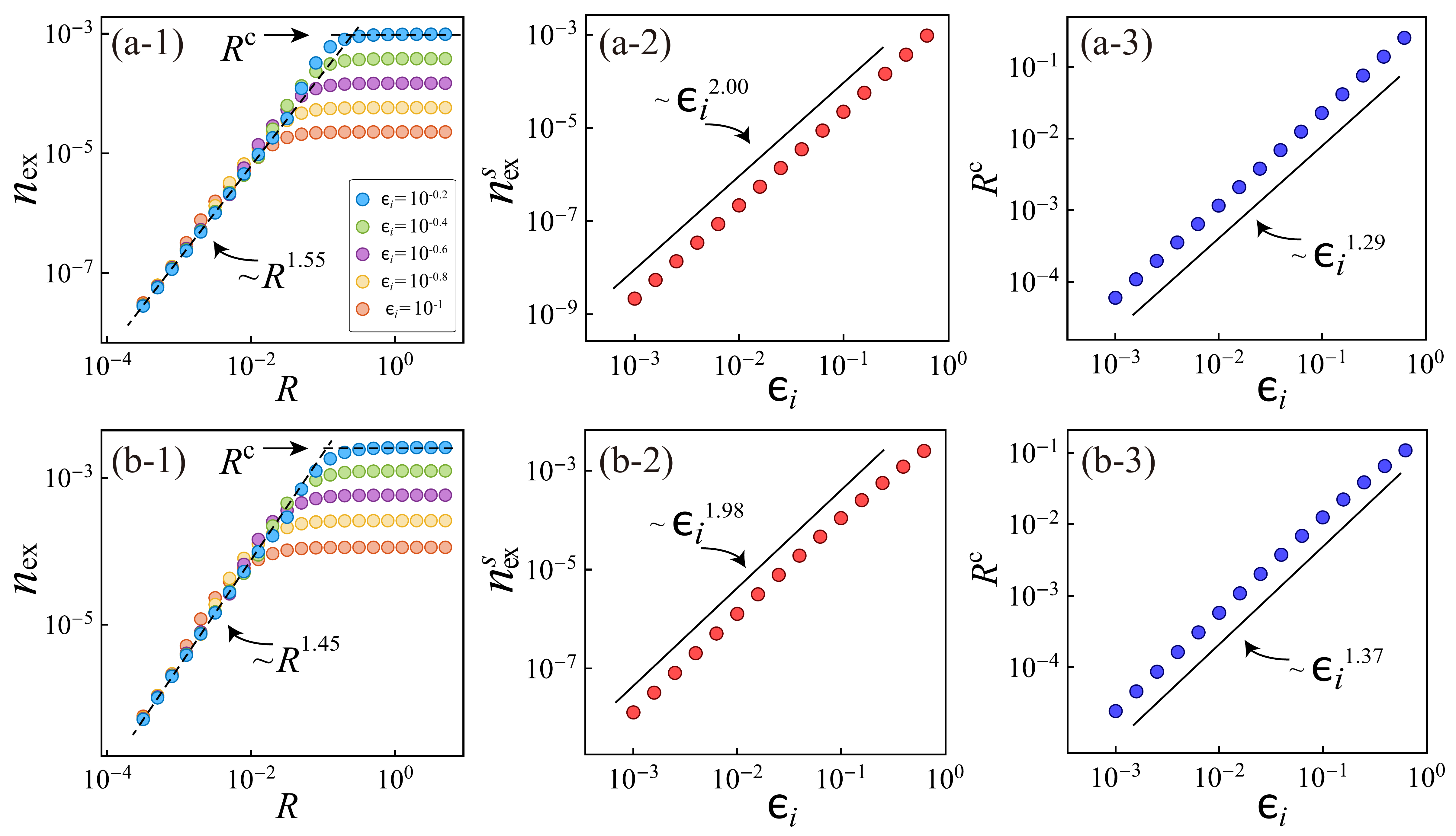}\\
	\caption{The anomalous dynamical scaling behaviors of the edge excitation number \( n_{\text{ex}} \) when the system is driven to topologically nontrivial QCPs with (a) four- and (b) six-fold degenerate edge modes are presented. (1) The dependence of \( n_{\text{ex}} \) on the quench rate \( R \) is illustrated for various values of \( \epsilon_i = 10^{-0.2}, 10^{-0.3}, 10^{-0.4}, 10^{-0.5}, 10^{-0.6} \) (from top to bottom). In the slow-quench regime, \( n_{\text{ex}} \) follows a power-law scaling with an exponent close to \( 1.5 \). In the fast-quench regime, the number of edge excitations  saturates to a value that is independent of the quench rate. (2) The saturation value  \( n^s_{\text{ex}} \) and (3) the critical quench rate \( R^c \) both exhibit power-law scaling with the dimensionless distance \( \epsilon_i \), with exponents close to \( 2 \) and \( \frac{4}{3} \), respectively. All plots are displayed on log-log scales.
} \label{fig:s4}
\end{figure}

In this section, we investigate universal dynamical scaling at QCPs with different topological degeneracies. To this end, we consider a generic Hamiltonian $ H = \sum_{\alpha} g_{\alpha} H_{\alpha} $. We first focus on a QCP hosting fourfold-degenerate edge modes $( \alpha = 0,1,2,3 )$. Fixing $ g_2 = 1 $, we linearly ramp $ g_3 $ according to $ g_3(t) = g_{3,i} + (g_{3,c} - g_{3,i})Rt $, with $ t \in [0, 1/R] $, thereby driving the system to a topologically nontrivial QCP at $ g_{3,c} = 1 $. Small couplings $ g_0 = g_1 = 0.05 $ are introduced to weakly couple the edge modes to bulk degrees of freedom and activate dynamical scaling. As shown in Fig.~\ref{fig:s4} (a-1), in the adiabatic regime of small $ R $, the edge-excitation number exhibits a power-law dependence, $ n_{\mathrm{ex}} \propto R^{1.55} $, with an exponent close to $ 1.5 $. Furthermore, Fig.~\ref{fig:s4} (a-2) and \ref{fig:s4} (a-3) show that the saturated excitation  number  $ n_{\mathrm{ex}}^{s} $ and the critical quench rate $ R^{c} $ obey universal power-law scaling with the initial distance to criticality $ \epsilon_i = (g_{3,i} - g_{3,c})/g_{3,c} $, with exponents $ 2.00 $ and $ 1.29 $, respectively, in close agreement with those reported in the main text.

Following a similar dynamical protocol, we fix $ g_3 = 1 $ and linearly ramp $ g_4 $ according to $ g_4(t) = g_{4,i} + (g_{4,c} - g_{4,i})Rt $, with $ t \in [0, 1/R] $, thereby driving the system to a topologically nontrivial QCP with sixfold-degenerate edge modes $( \alpha = 0,1,2,3,4 )$ at $ g_{4,c} = 0.95 $. As before, we also introduce small couplings $ g_0 = g_1 = g_2 = 0.05 $ to weakly couple the edge modes to the bulk and thus activate dynamical scaling. As shown in Fig.~\ref{fig:s4} (b-1), in the adiabatic regime, the  number of excitations  exhibits a power-law dependence on the quench rate with an exponent $ \approx 1.45 $, close to the value $ 1.5 $ reported in the main text. Moreover, in the fast-quench regime, both the saturated excitation number $ n_{\mathrm{ex}}^{s} $ and the critical quench rate $ R^{c} $ display power-law scaling with respect to the initial distance to criticality $ \epsilon_i = (g_{4,i} - g_{4,c})/g_{4,c} $, with exponents $ 1.98 $ and $ 1.37 $, respectively, as shown in Figs.~\ref{fig:s4} (b-2) and \ref{fig:s4} (b-3). These results are also consistent with those presented in the main text. 

Therefore, we conclude that the anomalous dynamical scaling behaviors reported in the main text emerge in a broad class of topologically nontrivial quantum critical points. More importantly, the associated anomalous dynamical power-law exponents are independent of the number of topological edge modes at criticality and are hence universal.





\begin{figure}[t]
	\centering
\includegraphics[width=3.5in]{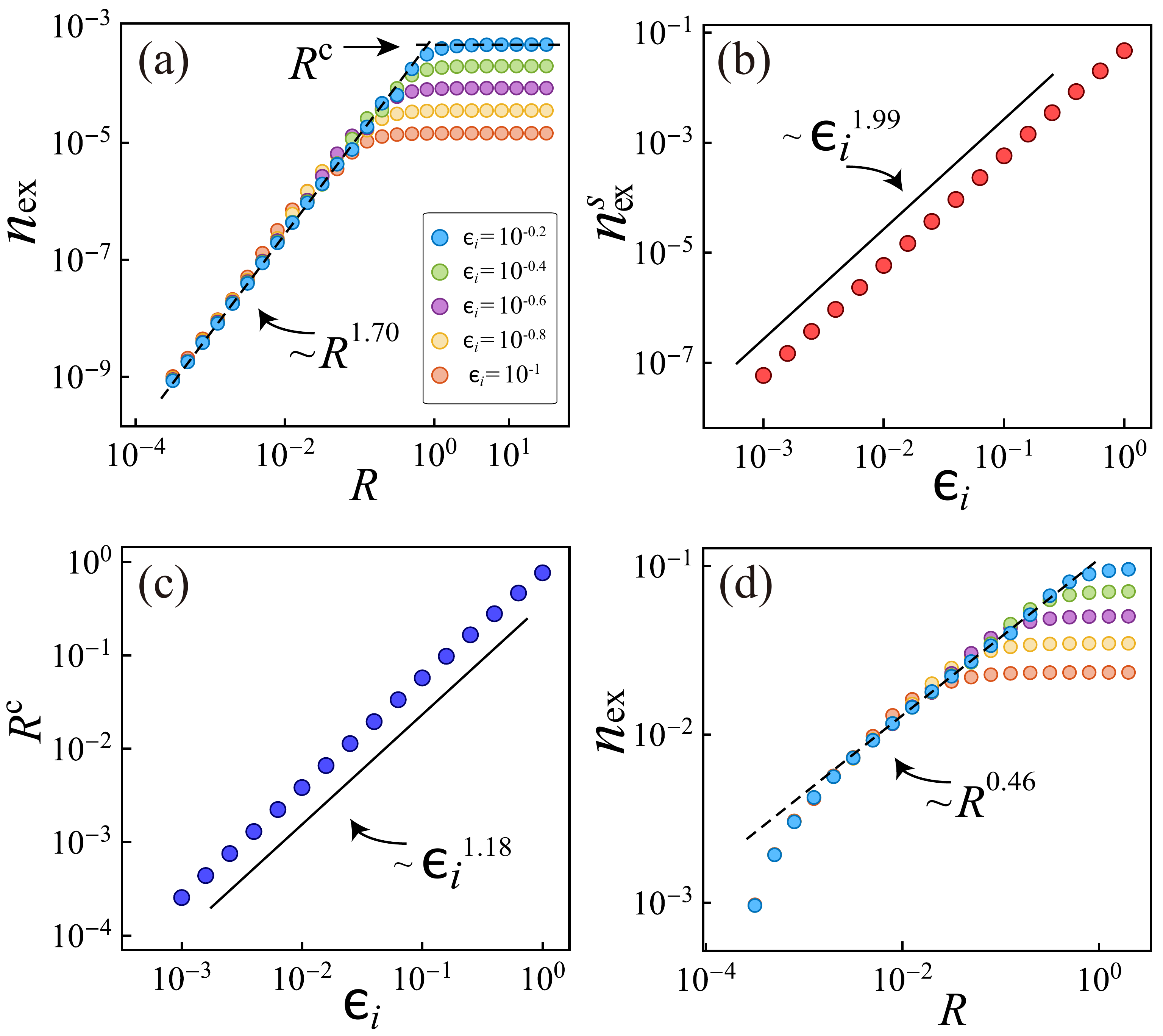}\\
	\caption{(a) Edge-excitation  $ n_{\text{ex}}$ as a function of the quench rate $ R $ for different initial distances from criticality, $ \epsilon_i = 10^{-0.2}, 10^{-0.3}, 10^{-0.4}, 10^{-0.5}, 10^{-0.6} $ (top to bottom). In the slow-quench regime, $ n_{\text{ex}} $ exhibits anomalous power-law scaling with an exponent $\simeq 1.7$. In the fast-quench regime, $ n_{\text{ex}} $ saturates to a value independent of $ R $.
(b) Saturated excitation number $ n_{\text{ex}}^{s} $ and (c) critical quench rate $ R^c $ as functions of $ \epsilon_i $, both displaying power-law scaling with exponents close to $2$ and $1.2$, respectively.
(d) Edge-excitation  $ n_{\text{ex}} $ generated by quenching to a topologically trivial QCP for different $ \epsilon_i $. In this case, $ n_{\text{ex}} $ follows the standard Kibble–Zurek scaling with an exponent close to 0.5~\cite{Ulifmmode2019PRB}. All data are shown on log–log scales.}
 \label{fig:s7}
\end{figure}

\section{Anomalous dynamical scaling and beyond KZ mechanism in two-dimensional topological nontrivial QCPs}
\label{sec:4}

In this section, we introduce a lattice model for Chern-insulator transitions that hosts nontrivial chiral edge states, enabling the exploration of anomalous driven dynamics in two dimensions.

Analogous to the one-dimensional case, we construct a two-dimensional (2D) lattice model exhibiting topological quantum criticality by forming linear combinations of fixed-point Hamiltonians with $\alpha$-range couplings on a square lattice. In momentum space, the model is defined as
\begin{equation}
\label{H_2d}
H_\alpha = \sum_{\bm k} \left( c_{\bm k,A}^\dagger, c_{\bm k,B}^\dagger \right) \mathcal H_\alpha(\bm k)
\begin{pmatrix}
c_{\bm k,A} \
c_{\bm k,B}
\end{pmatrix},
\qquad
\mathcal H_\alpha(\bm k) = \sin(\alpha k_x)\sigma^x - \sin(k_y)\sigma^y + \bigl[1-\cos(\alpha k_x)-\cos(k_y)\bigr]\sigma^z,
\end{equation}
where $A$ and $B$ label the two orbital (or sublattice) degrees of freedom within each unit cell. We then interpolate among $H_0$, $H_1$, and $H_2$ via:
\begin{equation}
H = \frac{1}{a+b+c}\left(a H_0 + b H_1 + c H_2\right), \qquad a,b,c>0 .
\label{eq:Chernmodel}
\end{equation}
The resulting phase diagram closely parallels the one-dimensional case (as shown in Fig.~\ref{fig:s2} (a)) and consists of insulating phases characterized by Chern numbers $C=0,1,$ and $2$~\cite{verresen2020topologyedgestatessurvive}. Importantly, the continuous transitions between $C=1$ and $C=2$, as well as between $C=1$ and $C=0$, correspond to topologically distinct critical points. The former supports topologically protected chiral edge modes, whereas the latter does not~\cite{verresen2020topologyedgestatessurvive,guo2025generalizedlihaldanecorrespondencecritical}.  This finding motivates us to investigate whether anomalous topology-dependent dynamical scaling—established previously in one dimension—also emerges in two dimensions. 

In the following analysis, we consider the model under periodic boundary conditions along the $y$ direction and open boundary conditions along the $x$ direction. Given the periodic boundary conditions along the $y$ direction, it is convenient to use the basis $\vert k_y\rangle \otimes \vert x \rangle$, where $k_y \in [0, 2\pi)$ represents the Bloch wave vector and $x \in \{1,...,L_x\}$ denotes a lattice site in the $x$ direction. In this basis, the Hamiltonian becomes block diagonal: $H = \sum_{k_y} \vert k_y\rangle\langle k_y\vert \otimes H(k_y),$ where 
\begin{equation}
\label{eq:hkx}
\begin{split}
H(k_y) &= \left( \sum_{x=1}^{L_x-1} \vert x\rangle\langle x+1\vert + \sum_{x=1}^{L_x-2} \vert x\rangle\langle x+2\vert \right) \otimes (b+c)\tau_3 + \text{H.c.} \\
&\quad + \sum_{x=1}^{L_x} \vert x\rangle\langle x\vert \otimes \{ a\tau_1 + (b+c)\tau_2 \},
\end{split}
\end{equation}
with $\tau_1 = -\cos k_y \sigma_z - \sin k_y \sigma_y$, $\tau_2 = -\cos k_y \sigma_z - \sin k_y \sigma_y$, and $\tau_3 = -\frac{1}{2}(\sigma_z + \sigma_y)$, where the $\sigma$ are Pauli matrices. An electron occupying the $m$th subband with eigenenergy $\varepsilon_m(k_y)$ is described by the wave function $\vert\Psi_m(k_y)\rangle = \vert k_y\rangle \otimes \vert u_m(k_y)\rangle$, where $\vert u_m(k_y)\rangle$ is an eigenstate of $H(k_y)$.

For the setup of the driven dynamics at 2D criticality, we fix $a(c)=0.2$ and $b+c(a)=3.8$, and ramp $b$ linearly as $b(t) = b + (b_c-b)Rt$, where $t \in [0, 1/R]$, at topologically trivial (nontrivial) quantum critical points, respectively, with $b_c=2.0$. As in the one-dimensional case, a small $a$ is included to activate dynamical scaling by weakly coupling to the bulk degrees of freedom. For the 2D system, the number of edge excitations is given by $n_{ex} = \frac{1}{L_y} \sum_{k_y} n_{ex}(k_y)$, where $L_y$ is the size along the $y$ direction and $n_{ex}(k_y)$ represents the momentum distributions of edge excitations, which is defined as~\cite{Liou2018PRB,Ulifmmode2019PRB}:
\begin{equation}
n_{ex}(k_y) = \sum_{n=L,R} \sum_{m \in v} |\langle u_n(k_y)|\psi_m(k_y)\rangle|^2 \Theta(\varepsilon_n(k_y)),
\label{eq:2de}
\end{equation}
where $|u_n(k_y)$ is an eigenstate of the final Hamiltonian $H(k_y)$, the sum over $n$ runs over the edge states, \(\Theta\) is the Heaviside function, and $\{|\psi_m(k_y)|, m \in v\}$ is the set of states obtained by time-evolving the states that formed the valence band before the quench.

For the topologically nontrivial quantum critical point, the resulting edge-excitation  $n_{\text{ex}}$ as a function of the quench rate $R$ is shown in Fig.~\ref{fig:s7} (a). In the slow-quench regime, the numerical results clearly demonstrate an anomalous power-law scaling of $n_{\text{ex}} \propto R^{1.70}$. In the fast-quench regime, the universal features identified in one dimension persist (see Fig.~\ref{fig:s7} (b) and (c)): the saturation value scales as $n^s_{\text{ex}} \propto \epsilon_i^{1.99}$, and the critical quench rate follows $R^c \propto \epsilon_i^{1.18}$. In contrast, at topologically trivial quantum critical points, the edge-excitation  approximately follows the standard KZ scaling, $n_{\text{ex}} \sim R^{0.5}$, as shown in Fig.~\ref{fig:s7} (d). These results demonstrate that the anomalous topology-driven dynamics at criticality are universal in two dimensions and cannot be illustrated through the conventional KZ mechanism, in sharp contrast to ordinary 2D quantum critical points.

\end{document}